\documentclass[letterpaper,twocolumn,10pt]{article}
\usepackage{usenix2019_v3}
\usepackage{tikz}
\usepackage{amsmath}
\usepackage{enumerate}
\usepackage{fancyhdr}
\usepackage{float}
\usepackage{stfloats}
\fnbelowfloat
\usepackage{url}
 
\usepackage{color}
\usepackage{booktabs}
\usepackage{multirow}
\usepackage{threeparttable}
\usepackage{environ}
\usepackage{amsmath} 
\usepackage{amssymb,mathrsfs,amsmath}
\usepackage{framed}
\usepackage{setspace}

\usepackage{listings}
\definecolor{codegreen}{rgb}{0,0.6,0}
\definecolor{codegray}{rgb}{0.5,0.5,0.5}
\definecolor{codepurple}{rgb}{0.58,0,0.82}
\definecolor{backcolour}{rgb}{0.95,0.95,0.95}
\lstdefinestyle{mystyle}{
 backgroundcolor=\color{backcolour},   
 commentstyle=\color{codegreen},
 keywordstyle=\color{magenta},
 numberstyle=\tiny\color{codegray},
 stringstyle=\color{codepurple},
 basicstyle=\footnotesize\ttfamily,
 breakatwhitespace=false,         
 breaklines=true,                 
 captionpos=b,                    
 keepspaces=true,                
 numbers=none,  
 numbersep=2pt,                  
 showspaces=false,                
 showstringspaces=false,
 showtabs=false,                  
 tabsize=2,
 xleftmargin=\parindent,
}

\usepackage{subfigure}
\usepackage[ruled,linesnumbered]{algorithm2e}
\usepackage{appendix}
\usepackage{float}
\usepackage{seqsplit}
\usepackage{tabularx}
\usepackage{graphicx}

\usepackage{authblk}

\usepackage{soul}
\usepackage{xcolor}

\newcommand{\code}[1]{{\ttfamily\seqsplit{#1}}}
  
\newcommand{\wqy}[1] {{\color{yellow!5!blue}#1}}

\newcommand{\ignore}[1]{}

\usepackage{etoolbox}
\newcommand{\system}{{\texttt{MPInspector}}\xspace}

\newcommand\blfootnote[1]{%
\begingroup
\renewcommand\thefootnote{}\footnote{#1}%
\addtocounter{footnote}{-1}%
\endgroup
}

\begin{document}

\title{\Large \bf MPInspector: A Systematic and Automatic Approach \\ for Evaluating the Security of IoT Messaging Protocols}

\pagestyle{empty}  

\author[$\dag$]{Qinying Wang}
\author[$\dag, \P$]{Shouling Ji}
\author[$+$]{Yuan Tian}
\author[$\dag, \P$]{Xuhong Zhang}
\author[$\S$]{Binbin Zhao} 
\author[$\dag$]{Yuhong Kan}
\author[$\dag$]{Zhaowei Lin}
\author[$\dag, \P$]{Changting Lin}
\author[$\dag, \P$]{Shuiguang Deng}
\author[$\ddag$]{Alex X. Liu }
\author[$\S$]{Raheem Beyah}

\affil[ ]{$^{\dag}${\small  Zhejiang University}, 
$^\P${\small Binjiang Institute of Zhejiang University}, 
$^+${\small University of Virginia},
$^\S${\small Georgia Institute of Technology},
$^\ddag${\small Ant Group}
}



\affil[ ]{{\small E-mails: ~~\{wangqinying, sji\}@zju.edu.cn,
~~yuant@virginia.edu, ~~zhangxuhong@zju.edu.cn, ~~binbin.zhao@gatech.edu, ~\{kan\_yuhong, leon.linzw\}@zju.edu.cn, ~~linchangting@gmail.com, ~~dengsg@zju.edu.cn, ~~alexliu@antgroup.com, ~~rbeyah@gatech.edu.}}



%
\maketitle



\begin{abstract}
Facilitated by \textit{messaging protocols} (MP), many home devices are connected to the Internet, bringing convenience and accessibility to customers.
However, most deployed MPs on IoT platforms are fragmented, which are not implemented carefully to support secure communication.
To the best of our knowledge, there is no systematic solution to perform automatic security checks on MP implementations yet.

To bridge the gap, we present \texttt{MPInspector}, the first automatic and systematic solution for vetting the security of MP implementations.
\texttt{MPInspector} combines model learning with formal analysis and operates in three stages: (a) using parameter semantics extraction and interaction logic extraction to automatically infer the state machine of an MP implementation, (b) generating security properties based on meta properties and the state machine, and (c) applying automatic property based formal verification to identify property violations.
We evaluate \texttt{MPInspector} on three popular MPs, including MQTT, CoAP and AMQP, implemented on nine leading IoT platforms. It identifies 252 property violations, leveraging which we further identify eleven types of attacks under two realistic attack scenarios.
In addition, we demonstrate that \texttt{MPInspector} is lightweight (the average overhead of end-to-end analysis is \textasciitilde4.5 hours) and effective with a precision of 100\% in identifying property violations.
\end{abstract}
\blfootnote{
Shouling Ji and Xuhong Zhang are the  co-corresponding authors.}

\section{Introduction}
Messaging protocol (MP) is critical for IoT platforms, as it connects IoT devices to the Internet and enables the communication between IoT devices, users, manufactures, and IoT app servers. 
IoT platforms offer customized MP implementations with different security schemes for IoT vendors.
For example, Google IoT Core adopts Json Web Token (JWT) for authentication \cite{jwt}. 
Unfortunately, MPs are hard to design correctly and several implementation flaws have been identified through ad-hoc manual analysis~\cite{mcateer2017security}.
These flaws lead to critical consequences, such as denial of service (DoS), sensitive data theft and malicious message injection \cite{kim2017automated, zhou2019discovering}.
So far, IoT platforms still have limited understanding about the security of MPs, since neither industry nor academia has good ways to systemically and effectively evaluate the security of MP implementations.
Considering the large amount of diversified IoT platforms, manual analysis that requires significant expert efforts is infeasible. 
Consequently, the pressing question is how to build an automatic tool to verify the security properties of MP implementations on different IoT platforms effectively? To answer the question, there are two main challenges.

\noindent\textbf{Diverse and customized MP implementations. } The MP implementations are diverse. 
Specifically, there are multiple types of MPs with different message formats and mechanisms, such as MQTT (Message Queuing Telemetry Transport) \cite{mqttstandards}, CoAP (Constrained Application Protocol) \cite{coap} and AMQP (Advanced Message Queuing Protocol) \cite{amqp}.
In addition, there are various customized implementations on different IoT platforms with different programming languages for each MP.
These diverse and customized MP implementations stress the scalability of the analysis.
Even worse, there are always gaps between the customized MP implementations and the standard MP specification, such as the differences on the configuration, parameter semantics, and interaction logic. 
Therefore, previous work on analyzing the high-level protocol specifications \cite{bhargavan2017verified, cremers2017comprehensive, hussain2018lteinspector, cremers2019component} is hardly applicable in the IoT context.

\noindent\textbf{Complex and closed-source MP workflow. }
Checking the MP implementation requires precisely modeling MP workflow including the exchanged parameters and interaction logic.
However, the workflow of MP is complicated, as it connects multiple devices and usually consists of multiple messages.
Even worse, MP implementations are closed-source.
As an example, the commercial platforms such as AWS IoT Core \cite{awsiotcore} and Azure IoT Hub \cite{azureiothub} do not open their source code on the server side. 
The closed-source MP implementation requires any testing approach to be black-box and system-agnostic.
Accordingly, previous research on program analysis for protocols \cite{lie2001simple, liu2021tifs, celik2018sensitive} cannot be used.

To handle these challenges, previous research conducts reverse engineering on the firmware and apps \cite{jia2020burglars}, which requires large expert knowledge.
Therefore, it is not scalable and can be time-consuming.
Fuzzing is an alternative solution \cite{li2020v, unifuzz, moptfuzz} to detect flaws by monitoring the crashes of the system under test.
However, it can hardly cover the full workflow of an MP implementation and cannot discover logic flaws that do not cause crashes.

\noindent\textbf{Our solution.}  To address the above challenges, we propose and implement \system, the first framework to systematically and automatically identify security flaws in MP implementations. 
We follow a property-driven and model-based testing philosophy.
First, we model an MP implementation into a state machine.
Second, we gather the security properties that need to be verified from the standard MP specification and refine them based on the learned state machine.
Finally, we detect property violations on the state machine by formal verification.
Specifically, the extracted state machine includes transition messages and transition logic. 
Transition messages are the messages that trigger the transition from one state to another, while transition logic is also referred to as interaction logic.
To support in-depth inspection of security flaws in MP implementations, \system recovers the detailed semantics of transition messages, which refer to as the customized composition of each parameter in the messages.
For example, the \texttt{ClientID} parameter in MQTT \cite{mqttstandards} may consist of \texttt{ProjectId} and \texttt{DeviceId} in a customized MP implementation.
As for the interaction logic, we adopt active model learning \cite{angluin1987learning}, a framework to construct the state machine of a system by providing inputs and observing outputs.
In \system, the inputs are messages sent to an MP implementation and the outputs are the relevant response messages or the connection states. 
Then, \system gathers security properties that need to be verified, which include the meta properties concluded from the standard MP specification and the extended properties inferred from the customized MP implementation.
After that, we convert the state machine and security properties into Tamarin codes and perform formal verification with Tamarin Prover\cite{tamarinmanual}. In the above procedures, we meet several challenges as follows. 

First, extracting message semantics is non-trivial, as some parameters may be encrypted, making their semantics hidden.
To tackle this, we construct traffic- and NLP-based methods to identify the crypto function of each encrypted parameter.
Then, the semantics of a parameter can be recovered according to the definition of the identified crypto function.
Some common crypto functions can be identified by pattern matching on the real traffic, while it is almost impossible to define patterns for the unknown customized crypto functions.
Since the parameters with customized crypto functions are usually specified in the IoT manufacturer documents offered by IoT platforms, we further develop a novel NLP-based method to directly extract the semantics of these parameters from the IoT manufacturer documents.

Second, considering the IoT context that involves multiple parties and multiple types of messages, active model learning cannot be directly applied to extract the interaction logic of MP implementations, as it only supports two parties and can be time-consuming when dealing with multiple types of messages.
Moreover, when applying model learning to test MP implementations in the real world, they may produce uncertain responses due to uncontrolled factors, e.g., failing to receive an expected response due to timeout.
In such a case, model learning may be trapped into an endless learning procedure, thereby failing to construct the state machine.
To overcome these issues, we design an enhanced active model learning framework to support observing outputs from multiple parties.
Further, to speed up the learning procedure, \system cuts down unnecessary input tests. To overcome the uncertainty issue, \system stops the learning procedure if the same state machine is constructed more than once.

Third, when performing formal verification, the traditional Tamarin Prover may fail to prove some properties, as some MP implementations have complex state transitions.
In order to solve this problem, we design a helping oracle to guide the proof, which is a script that can help Tamarin Prover adjust the order of solving goals during the proof.

\noindent\textbf{Evaluation.} We apply \texttt{MPInspector} on three popular MPs, MQTT, CoAP and AMQP, implemented on nine leading IoT platforms (e.g., Google IoT Core, Azure IoT Hub) \cite{al2015internet}. It successfully recovers the state machines of all the MP implementations and formally verifies their authentication and secrecy properties. The average overhead of end-to-end analysis is 4.5 hours with a precision of 100\% in identifying property violations. Specifically, it checks 57 customized security properties and detects 252 property violations, leveraging which we further identify eleven types of attacks. 
These results and findings are alarming. Each platform at least violates 18 properties, which enables at least one attack. The resulting attacks have serious consequences, e.g., privacy leakage and malicious data injection.
Our research further shows that the main root causes of risky MP implementations are: (1) the gap between ad-hoc MP implementations and the standard specification, (2) the undermined security mechanisms under the resource constrained IoT context, and (3) the lack of careful consideration about device sharing, multi-party involved communication situations under the IoT context.

\noindent \textbf{Summary and contributions.} Our key contributions are:
\begin{itemize}
\item We propose \system, the first framework for automatic security analysis of MP implementations. \system is precise on the detection of MP implementation flaws and is extensible and configurable to different IoT platforms and different protocols.
We release \system as an open-source tool for facilitating further studies. 
\item With \system, we evaluate three popular MPs on nine leading IoT platforms and detect 252 property violations. We also uncover eleven kinds of attacks that exploit the combinations of property violations under practical threat models. We have responsibly reported these vulnerable implementations to the vendors and got acknowledged from vendors such as Tuya Smart.
\end{itemize}

\vspace{-3mm}
\section{Background}
\vspace{-2mm}
\subsection{Cloud based IoT Platforms} \label{sec:cloudbasediotplatforms}

\begin{figure}[hbtp]
  \centering
  \includegraphics[width=0.48\textwidth]{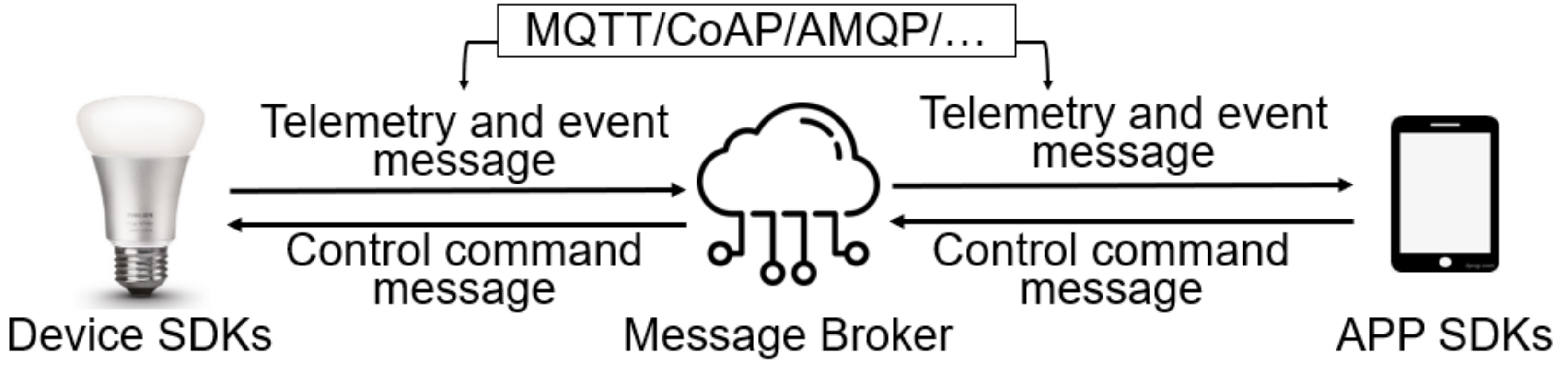}
  \vspace{-5mm}
  \caption{A typical architecture of MP implementations.}
  \label{mp} 
\end{figure}

Today, most IoT platforms (e.g., AWS and Azure) offer MP implementations, which serve as networking infrastructures for IoT manufactures and also called SaaS (Software-as-a-Service) applications. 
As shown in Figure~\ref{mp}, the service contains the message broker (can be configured by IoT manufactures), device SDKs (e.g., cameras and lockers) and APP SDKs (designed for terminal users). 
The device sends telemetry and event messages and receives command messages via MPs, and the user application also sends control commands to the devices remotely via MPs. We regard the device and the application as clients.
All the messages between the device and the application are forwarded by the broker on the remote IoT platform. We regard the broker as the server.
IoT device manufactures buy and deploy the SaaS application for MP to enable users remotely control their devices.

Studying the SaaS appliactions for MPs can cover most devices in the real world. A previous survey \cite{whyshouldbuildown} shows that IoT manufactures simply deploy the SaaS without customization.
As a result, security analysis of the SaaS appalications for MP can reflect the real-world threats.

\vspace{-3mm}
\subsection{MP Types and Implementations}
\label{sec:customizedmpimplementations}

Various MPs with distinct message types and formats have been implemented for IoT systems. 
For example, MQTT has nine key types of messages running over TCP.
Among them, CONNECT is one type of MQTT messages, and it has five key parameters including \texttt{ClientID, Username, Password, WillTopic} and \texttt{WillMessage}.
Meanwhile, CoAP has two types of messages running over UDP.
Among them, CON is one type of CoAP messages, and it has six key parameters including \texttt{Uri, MessageId, Request, Option, Token} and \texttt{Payload.}
For existing MPs, MQTT, CoAP and AMQP are the three most prominent MPs adopted by IoT platforms \cite{al2015internet}.
For more details and distinctions about these MPs, please refer to their standard specifications \cite{mqttstandards,coap,amqp}. 


Based on the standard MP specification, MP implementations can be customized by the IoT platforms, including the configuration, the parameters in the messages and the message interaction logic. 
As for configurations, IoT platforms such as Aliyun Cloud and Tuya Smart optionally adopt the secure session protocol such as SSL/TLS. The configuration of secure session protocol may also be customized by IoT platforms.
For example, Google IoT Core and Azure IoT Hub do not support authenticating a client by the certification on the server side. Instead, they adopt customized tokens for authentication.
As for parameters, the parameters in messages can have customized semantics. 
For example, on AWS IoT Core, the \texttt{Username} and \texttt{Password} are not adopted in the implementation, while on Google IoT core, \texttt{Username} in a CONNECT message is composed of \texttt{ProjectId} and \texttt{deviceId}, e.g., \texttt{light123/dev1}. Besides, Tuya Smart assigns a control command and a timestamp to the payload in the PUBLISH message and encrypts these values by a private key using a customized crypto function.
Moreover, the message interaction logic can be customized. As an example, Bosch IoT platform allows two clients with the same \texttt{ClientID} to be connected with the server at the same time, which is, however, not allowed in the standard MQTT specification.

 \vspace{-5mm}
\section{Threat Model} \label{sec:threatmodel}
\vspace{-3mm}
We consider two practical attack scenarios as follows.

\noindent\textbf{Neighbor scenario.} In this scenario, the victim and attacker are within the same local network, e.g., in rental homes, and the attacker can perform network-based exploits.
We apply the standard Dolev-Yao threat model \cite{dolev1983security} on the communication channel, under which the attacker can eavesdrop and modify all messages transferred on this channel and can impersonate a legitimate participator to inject messages.

\noindent\textbf{Tenant scenario.} Inspired by previous works \cite{janes2020never, jia2020burglars}, the tenant scenario characterizes the situations where a victim uses some devices previously used by an attacker.
Such cases include second-hand devices \cite{ebaystrongiot} and devices in hotels, Airbnb and rental homes \cite{dey2020exploring}.
In this scenario, when the attacker owns the device, he/she can collect the device identity including the password of the device or leave a backdoor on the device. 
After that, when the device is delivered to the victim, the attacker can use the collected identity or the injected backdoor to conduct attacks by sending some malicious command or publishing fake state of the device.

In both scenarios, the goal of the attacker is to exploit the flaws in the client-server interaction to take control of the victim device or monitor/manipulate the victim device data.

\section{Design and Implementation } %


\subsection{Overview}\label{sec:overview}
\begin{figure*}[htb]
  \centering
  \includegraphics[width=0.88\textwidth]{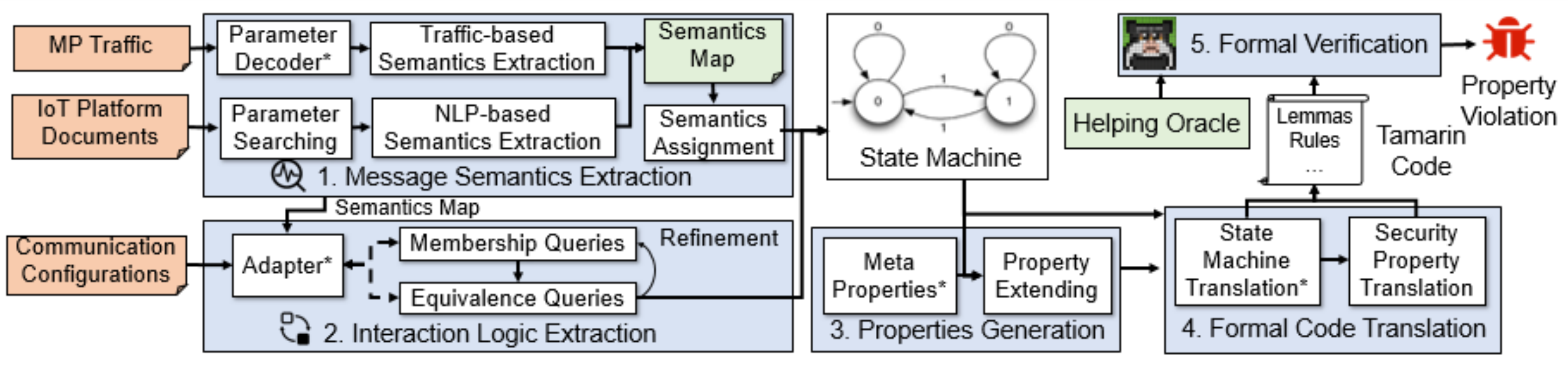} 
  \vspace{-3mm}
  \caption{Overview of \texttt{MPInspector}. \system supports automatically testing of any customized implementation of MQTT, CoAP, or AMQP out of the box. To support a new type of MP, the modules labeled with a star need to be extended. 
  }
  \vspace{-0.2in}
  \label{fig:mpinspector} 
\end{figure*}
At a high level, \system aims to automatically verify the security properties of MP implementations on different IoT platforms.
Figure~\ref{fig:mpinspector} provides an overview of \system, which includes five modules: message semantics extraction, interaction logic extraction, property generation, formal code translation and formal verification.

The workflow is as follows.
First, the message semantics extraction module accepts MP traffic and IoT platform documents as inputs, and extracts the customized composition semantics of each parameter specified in the standard MP specification.
Second, the interaction logic extraction module performs active model learning to infer the raw state machine by sending messages to the involved parties in the MP implementation and monitoring their responses. This module requires users to specify the communication configuration in order to generate the messages in the learning process.
After these two stages, \system adds the message semantics extracted from the first module to the transition messages in the raw state machine inferred in the second module to form a detailed state machine.
Third, the property generation module extends the meta properties from the standard MP specification with the extended properties inferred from the detailed state machine to form the final security properties to be validated.
Fourth, the formal code translation module translates the detailed state machine and security properties into Tamarin code.
Finally, \system applies Tamarin Prover to perform formal verification on the Tamarin code. The final outputs are the violated security properties.
To make a clearer clarification, we take the MQTT implementation on the Bosch IoT platform as a running example to explain the main process, which is shown in Appendix~\ref{app:arunningexample}.
\subsection{Inputs}
\system takes three inputs: MP traffic, IoT platform documents and communication configurations.

\noindent\textbf{MP traffic.} \system accepts MP traffic to extract message semantics. 
The analyst can collect the traffic using his/her device and application to interact with the broker.
He/she can set an access point (AP), to which his/her device and application are connected. Then, he/she can apply Wireshark or SSLSplit to record the traffic produced during the interaction.
To collect as many different types of messages as possible, the analyst can perform different actions on the client, including sending commands and changing the state of the client.

\noindent\textbf{IoT platform documents.} 
IoT platform documents are supplements to identify the semantics of parameters that cannot be identified from MP traffic.
IoT platforms generally offer rich semantics of these parameters in their publicly available documents for IoT manufacturers.
However, the downside of the semantics information in the documents is that it might not match the real implementation.
Therefore, we treat the documents as a secondary input and only use it when the parameter semantics cannot be extracted from the MP traffic.

 \noindent\textbf{Communication configurations.} These configurations are required for \system to generate real messages to communicate with the broker in the model learning process.
They include the MP type and key communication arguments of the device or application, which can be collected from the device's or application's configuration file.
Taking MQTT as an example, the key communication arguments are broker address, MQTT version, IoT platform name, raw password, secret key of the device or the application if exists, and the certifications if exist.


\subsection{Message Semantics Extraction}
\label{sec:parametersemanticsextraction}

The message semantics extraction module aims to extract the composition semantics of parameters in a message, which are of two types.
First, a parameter can be a composition of several terms concatenated with delimiters, e.g., parameter \texttt{Username} with value \texttt{light123/dev1} is composed of \texttt{ProjectId} and \texttt{DeviceId}.
Second, a parameter can be the encryption of several terms by a certain crypto function, e.g., a \texttt{Password} with the value of a complex character string can be the encryption of \texttt{ProjectId} and \texttt{ExpiredTime} by the JWT function \cite{jwt}.
Identifying the semantics of the second type of parameters is not trivial, as the value in the traffic does not have any meaning. 
To extract these two kinds of semantics, we provide two alternatives in this module.
As shown in Figure~\ref{fig:mpinspector}, the message semantics extraction module mainly consists of traffic- and NLP-based semantics extraction.
As the semantics extracted from the real MP traffic reflect the actual MP implementation, we prioritize the traffic-based semantics extraction.
For the parameters whose semantics cannot be identified from the MP traffic, we resort to the NLP-based semantics extraction.
Both of these methods output a semantics map, which maps the parameter values to their corresponding semantics.
For example, the pair \texttt{\{light123:ProjectId\}} means the semantics of the parameter \texttt{light123} is \texttt{ProjectId}.
In the last step, the two returned semantics maps are merged and fed to the semantics assignment component, which then replace the values in a message with the matched semantics from the semantics map.
For parameters having no match in the semantics map, we still need to assign each of them a specific name for the following modeling task.
Thus, we sequentially assign them a fake semantics, e.g., \texttt{V0}, \texttt{V1}, \texttt{V2}.
Taking the parameter \texttt{ClientID} as an example, its extracted semantics may look like \code{(V0, aud, V2, V3)} where the \texttt{aud} means audience.
Below, we detail the traffic- and NLP-based semantics extraction process.


For the traffic-based semantics extraction, the parameter parsing component first takes MP traffic as input and decodes the messages from the MP traffic to extract the values of the parameters.
For some parameter values, their semantics can be directly inferred from the traffic, e,g., the \texttt{Payload} in a \texttt{PUBLISH} message may contain the format as \code{key : value} or \code{key = value}, and we can directly extract the key as the semantics of the value. 
Besides, there are also encrypted parameters whose semantics can only be recovered by identifying the corresponding crypto function.
For common crypto functions, we find that the encrypted values have common patterns, e.g., the common pattern for JWT is \texttt{$ey[A-Za-z0-9\_\backslash\backslash/+-]*\backslash\backslash.[A-Za-z0-9.\_\backslash\backslash/+-]*$}).
In our implementation, we provide the patterns of nine common crypto functions (e.g., JWT function and Base64 encoding).
The semantics extracted from the aforementioned process are also added to the semantics map.

For the parameters whose semantics cannot be extracted from the MP traffic, e.g., the ones encrypted by unknown customized crypto functions, we propose an NLP-based semantics extraction method.
Specifically, it extracts the semantics from IoT platform documents, which generally specify the semantics of parameters. 

\begin{figure}[htb]
  \centering
  \includegraphics[width=0.5\textwidth]{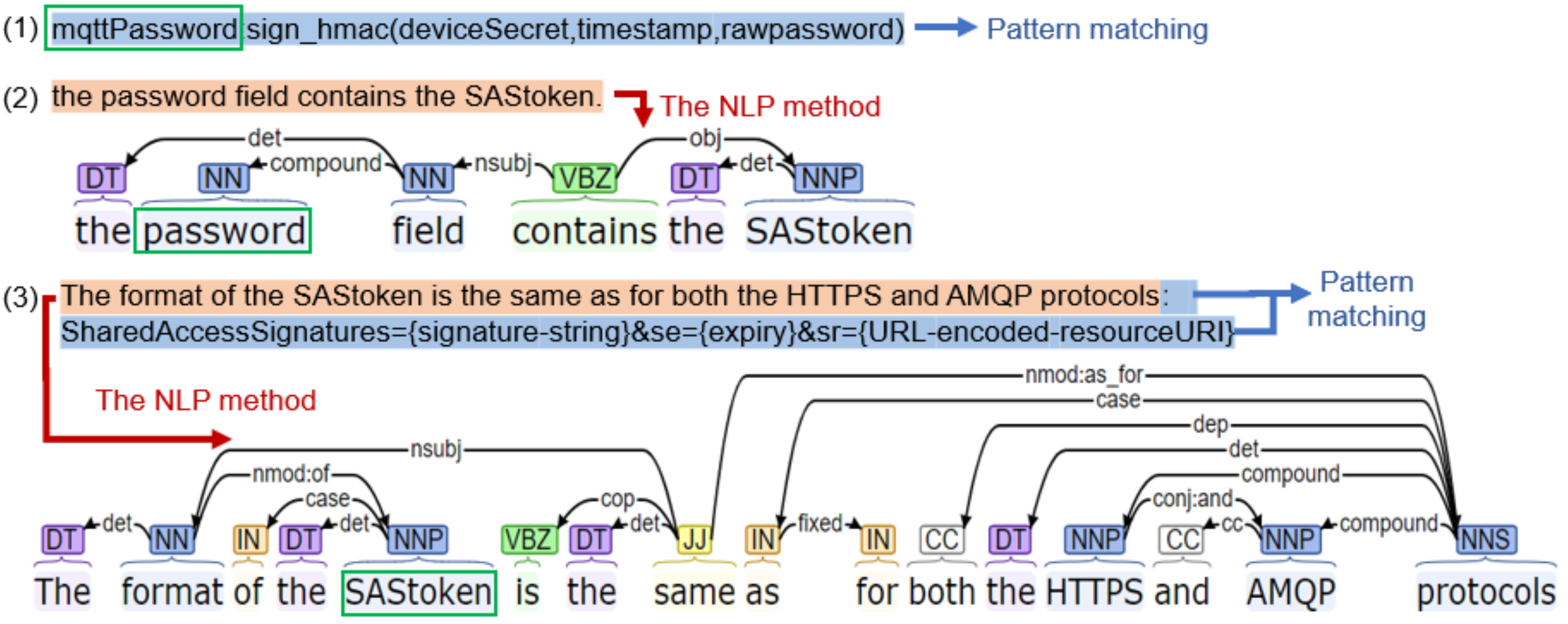} 
  \vspace{-5mm}
  \caption{Example sentences of three types, including the structured, unstructured, mixed sentences.
  }
  \label{fig:nlpmethod} 
\end{figure}

However, IoT platform documents are usually loosely formatted with sentences in different formats, posing challenges to semantics extraction.
In our observation, the documents mainly include three types of sentences  as shown in Figure~\ref{fig:nlpmethod}: (1) structured sentence;
(2) unstructured sentence in natural language;
and (3) a mixed type sentence that contains both structured and unstructured parts.

Based on the above observation, we take the following steps.
The parameter searching component takes IoT platform documents as input and parses sentences from the documents. 
For each parameter whose semantics cannot be extracted from the MP traffic, this component searches the sentences that contain the parameter.
Then, the NLP-based semantics extraction component divides the sentences into the above three types and analyzes the three types of sentences one by one.
This component first tries to extract semantics from the structured sentences.
If not success, it extracts semantics from the mixed sentences and finally the unstructured sentences.
The identified semantics will also be stored into a similar semantics map that will be used in the final semantics assignment component.

In detail, for structured sentences, they have obvious structure and symbols that indicate the parameter semantics, which can be extracted by pattern matching. 
For unstructured sentences, the idea is to find a noun or a noun phrase that has an equivalence or inclusion relation with the target parameter.
Thus, this module applies the Stanford dependency parser \cite{manning2014stanford} to identify the equivalence relation
and Part-of-Speech tagger \cite{manning2014stanford} to identify the part of speech of each word in the sentence.
For example, for the unstructured example in Figure~\ref{fig:nlpmethod}, we can identify the target parameter \texttt{password} has the inclusion relation with the \texttt{SAStoken}, indicated by the word \texttt{contain}.
For mixed sentences, the idea is to find the sentences satisfying two conditions: (1) the subject of the unstructured part is the target parameter, and (2) the structured and the unstructured parts are connected by equivalence symbols such as \texttt{:} and \texttt{=}, which indicate they have equivalence relation.
Finally, this component performs pattern matching on the structured part to extract the semantics.
For the mixed sentence example in Figure~\ref{fig:nlpmethod}, \texttt{MPInspector} first divides the sentence into a structured part in blue and an unstructured part in yellow by the delimiter \texttt{:}. Then \texttt{MPInspector} identifies that the subject of the unstructured part is composed of the target parameter \texttt{SAStoken}, and finally applies the pattern matching to the structured part to identify the semantics of \texttt{SAStoken}.



\subsection{Interaction Logic Extraction}
\label{sec:interactionlogicextraction}
This module aims to extract the raw state machine of the MP broker, since it is responsible for processing messages from clients and is closed-source.
The state machine includes transition messages and transition logic. 
Transition messages represent the messages that are used to trigger the transition from one state to another, consisting of the input message to the broker and the response message from the broker.
This module adopts active model learning, a framework to construct the state machine of a system by providing inputs and observing outputs.
In \texttt{MPInspector}, the inputs are different permutations of message sequences sent to the MP broker and the outputs are the relevant response message sequences.

The basic model learning procedure is as follows. First, this approach adopts membership queries (MQs) to collect the responses to the inputs, and generates a state machine (also noted as a hypothesis).
Then it performs equivalence queries (EQs) to seek an input that makes the hypothesis state machine and the real system have different outputs.
This input is also called a counterexample that distinguishes the inferred state machine and the real system.
If there is no counterexample, the inferred state machine is equivalent to the real system and is the final output of the interaction logic extraction module.
Otherwise, a new round learning with MQs and EQs will be performed until there is no counterexample.

As shown in Figure~\ref{fig:mpinspector}, we have three components in this module: adapter, MQ and EQ.
The adapter is designed to generate different input messages, send input messages to the broker, collect the response messages from the broker, and decode the response messages to identify their types.
When generating an input message, the adapter directly uses the parameter values from the semantics map in Section~\ref{sec:parametersemanticsextraction}. 
However, some parameters have dynamic values, e.g., a timestamp, which need to be generated by referring to their semantics in the semantics map.
In addition, there are some dynamic parameters that are encrypted, for which the adapter follows the cryptographic algorithm in the their semantics to generate their values.
Specifically, the adapter invokes the corresponding pre-installed encryption interface in \texttt{MPInspector}.
For example, for \texttt{mqttPassword} introduced in Figure~\ref{fig:nlpmethod} from Section~\ref{sec:parametersemanticsextraction}, the adapter invokes the HMAC interface and performs encryption of the timestamp and the raw password to generate the value of the parameter \texttt{mqttPassword}.

We implement the adapter for MQTT, CoAP, and AMQP, respectively. Based on the inputs and responses, MQs and EQs can infer the state machine of the broker.

The adapter in existing model learning frameworks usually only supports the communication of two parties, which is not applicable in the IoT context where multiple parties are usually involved.
To tackle this, we extend the adapter by the following steps: (1) extending the adapter to support sending all types of messages that can be sent to the broker from all clients, and (2) monitoring the responses of the broker and all clients.
Also, there are implicit responses from the broker.
For example, in MQTT, the broker may accept the input message but give no response. In addition, the broker may accidentally close the connection without sending any response message.
Therefore, we further extend the adapter to monitor the connection state of the broker and map the above two situations to two responses: \texttt{EMPTY} and \texttt{CONNECTIONCLOSED}, respectively.

Considering there may be many types of messages in the IoT context, the EQ component of existing model learning frameworks, e.g., Chow's W-Method \cite{Chow1995Testing}, needs to send message sequences for all the permutations of the message types to the broker, leading to a high performance overhead.
Therefore, we design a customized EQ component inspired by the previous work \cite{de2015protocol} to avoid useless queries to improve the efficiency.
Specifically, we add a check to see if the connection has been closed when testing a sequence of input messages.
If so, our learning procedure stops seeking counterexamples with this particular prefix of message sequences, as the following message sequences with this prefix will receive the same response, namely \texttt{CONNECTIONCLOSED}.
Thus, it does not make sense to continue searching for counterexamples with this prefix.
Our experiments prove that the customized EQ component reduces the query time by 34\% compared to Chow's W-Method.

\begin{figure}[!htb]
    \centering
    \includegraphics[width=0.45\textwidth]{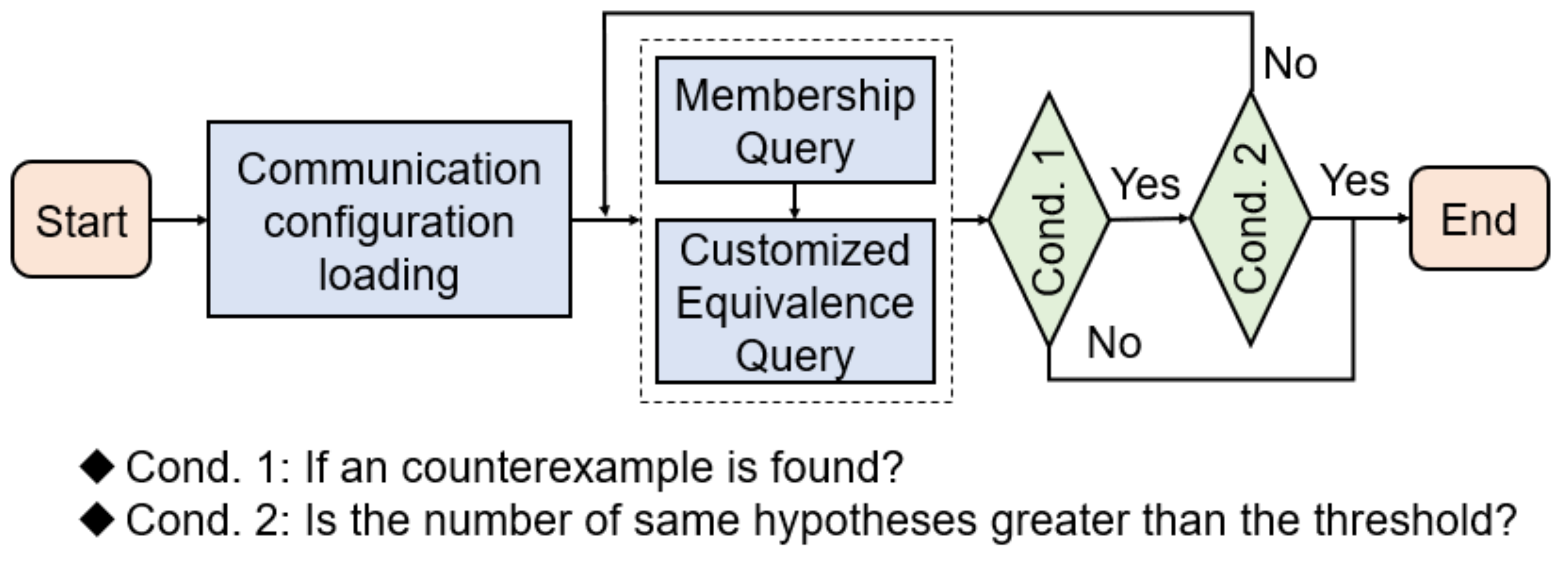}
    \vspace{-6mm}
    \caption{The learning procedure of active model learning.
    }
    \label{fig:modellearningstrategy}
\end{figure}

Another challenge is that existing active learning models may be trapped into an endless learning procedure and thus fails to construct the state machine. 
For instance, when applying model learning in the real world, the targeted broker may produce uncertain responses, e.g., \texttt{EMPTY} response caused by timeout, due to uncontrolled factors such as environment.
The EQ component may mistakenly take the uncertain response as a counterexample, which may further cause the same hypothesis to be generated repeatedly.
To tackle this, we observe that the same hypothesis is generated if and only if it is equivalent to the MP broker. 
Therefore, we limit the maximum amount of the same hypothesis that is generated repeatedly to help terminate the learning procedure, which is shown in Figure~\ref{fig:modellearningstrategy}.
Additionally, we set a time delay to wait for the broker's response for a query, which can mitigate the uncertain response issue when performing MQs and EQs.
The thresholds for the amount of the same hypothesis and the time delay can both be specified in the communication configurations.

After model learning, a raw state machine is generated whose transition messages only contain message names, e.g., \texttt{CONNECT/CONNACK}.
Then, \texttt{MPInspector} adds the message semantics extracted from Section~\ref{sec:parametersemanticsextraction} to the transition messages in the raw state machine.
In addition, we check if the MP implementation adopts SSL/TLS. 
If so, we insert the state transition with \code{KEYEXCHANGE\{sesson\_key\}} after the initial state to denote the SSL/TLS mechanism, and add the SSL/TLS encryption semantics on the transition messages.

Apart from the inferred state machine, some unobservable internal protocol states called validity predicates can not be extracted by the model learning method and need to be modeled in Section~\ref{sec:formalcodetranslation} for verification.
In our study, a validity predicate describes a constraint that a parameter should satisfy in a transition, e.g., the client's signature in a \texttt{password} parameter should be valid, or the current message ID should be less than the received message ID.
Thus, \texttt{MPInspector} extracts the validity predicates by utilizing the adapter to send messages with carefully mutated parameters to the server and observing if they are accepted or not.
Particularly, \texttt{MPInspector} supports extracting the validity predicates with the \texttt{Equality} and \texttt{LessThan} constraints.
Below are the corresponding mutation strategies.
For the parameter with numerical type, \texttt{MPInspector} mutates it by adding or subtracting a random number to it.
For other parameters, \texttt{MPInspector} changes one bit of their value for mutation.


\subsection{Property Generation} 
\label{sec:propertiesgeneration}
The property generation module generates the security properties that should be verified on the extracted state machine.
It aims to generate two groups of properties, including secrecy properties and authentication properties.
The secrecy properties are for the confidential goal of certain parameters and the authentication properties are used to check if certain types of messages are authenticated.
The parameters and messages that should be checked are first concluded from the standard MP specifications.
This initial set of security properties are also called meta properties, including the secrecy properties (e.g., \texttt{Meta\_Sec\_Set = \{ClietnID, Username, Password,...\}}) and the authentication properties (e.g., \texttt{Meta\_Auth\_Set = \{CONNECT, CONNACK, SUBSCRIBE,...\}}).
Second, we filter meta properties, whose targeted messages or parameters do not appear in the inferred state machine, as not all of the messages and parameters from the standard specification are used in IoT implementations.
Finally, we add the extended properties based on the inferred state machine, as messages of the same type may have different parameter semantics in an MP implementation.
For example, the CoAP implementation on Aliyun Cloud adopts two different \texttt{CON} messages with different parameter semantics for connecting and publishing messages to the broker, respectively.
Thus, we add the parameters from such different messages to the secrecy property set and such different messages to the authentication property set. 
In conclusion, the only hard-coded part in the property generation module is the meta properties from the standard MP specifications.
Note that this hard-code effort is required per MP type not per MP implementation.
We demonstrate the generated detailed security properties for MQTT, CoAP and AMQP in Appendix~\ref{app:properties}.

\subsection{Formal Code Translation} \label{sec:formalcodetranslation}
The formal code translation module aims to translate the inferred state machine and security properties into Tamarin code, which can be further analyzed by Tamarin Prover. There are two components in this module including state machine translation and security property translation.

The inferred state machine is translated into rules in Tamarin, where a rule defines a transition in the state machine.
A rule has a name and three parts, each of which is a sequence of facts: one for the rule’s left-hand side, one for the rule's middle part called action fact, and one for the rule’s right-hand side.
Taking the simplified transition messages \texttt{CONNECT/CONNACK} that trigger the broker from state \texttt{A} to state \code{B} as an example, the transition indicates the broker receives a \texttt {CONNECT} message in state \texttt{A}, which is modeled as two facts including the fact \texttt{In(connect)} and the fact \texttt{State\_A\_broker}.
The above two facts are put into the rule's left-hand side.
The transition indicates the broker turns into state \texttt{B} and sends out a \texttt{CONNACK} message, which is modeled as two facts including the fact \texttt{State\_B\_broker} and the fact \texttt{Out(connack)}.
The above two facts are put into the rule's right-hand side.
The action facts reason about the behaviours in the transition.
For example, we use \texttt{Commit(broker, connect)} to reason one of the behaviours of the transition \texttt{CONNECT/CONNACK}.
The rule supports let-binding expressions to specify the parameters in the message along with the detailed semantics, e.g., \texttt{connect = <a,b>}. After that, we have a simplified rule of the transition as shown in Listing~\ref{lst:basicrule}.


We translate the transition messages from the perspectives of both the broker and the client to completely model an MP implementation.
For example, \texttt{CONNECT/CONNACK} depicts the transition of the broker that it enters a new state and sends out a \texttt{CONNACK} message after receiving a \texttt{CONNECT} message.
It also depicts the two transitions of the client: one describes that the client enters state \texttt{D} from a former state \texttt{C} after sending a \texttt{CONNECT} message to the broker, and another describes that the client enters state \texttt{E} from state \texttt{D} after receiving a \texttt{CONNACK} message from the broker.

\begin{lstlisting}[breaklines,basicstyle=\footnotesize,caption={An example rule in Tamarin code.},label={lst:basicrule}, frame = single]
rule broker_recv_connect_snd_conncak:
  let
    connack = <a>
    connect = <a, b>
  in [ In(connect), State_A_broker ]
  --[ Create('connect',broker),
      Commit(broker,client,connect),
      Running(broker,client,connack) ]
    ->[ Out(connack), State_B_broker]
\end{lstlisting}

Additionally, for the validity predicates extracted from Section~\ref{sec:interactionlogicextraction}, \texttt{MPInspector} models them as a kind of action fact for the related rule's middle part.
Particularly, \texttt{MPInspector} adopts the kind of action fact called restriction, which is offered by Tamarin.
Restrictions specify constraints that a protocol transition should uphold, e.g., \texttt{Equal(x,y)} and \texttt{LessThan(x,y)}.
Since some validity predicates have the encryption semantics, \texttt{MPInspector} adds the corresponding encryption function to its action fact, e.g., \texttt{Equal(verify(sig,m,pubkey),true)}, where \texttt{verify(sig,m,pubkey)} is a predefined function in Tamarin to verify the signature \texttt{sig} on the received message or parameter \texttt{m}.
This action fact indicates that the \texttt{verify} function equals to the constant \texttt{true}.

When translating the state machine, we first implement the initialization rules based on the provided initial state to set up the initial parameters that the broker and clients own. 
The initialization rule has a sequence of facts that describe the initialization of parameters in its left-hand side and a sequence of facts that describe the initial state in its right-hand side.
Then, if the state machine considers the session key negotiation, we hard-code a general rule to model the transition, which is a simplified SSL/TLS key negotiation modeling.
Finally, we follow the above translation principle to translate the transition messages into rules. 

After state machine translation, the security properties are translated into lemmas, which are first-order logic formulas over time points and action facts, based on the standard security property templates specified from Tamarin Prover documentations \cite{tamarinmanual}.
Particularly, for each authentication property, \texttt{MPInspector} applies four types of authentication lemmas based on Lowe's taxonomy of authentication goals \cite{lowe1997hierarchy} to make a fine-grained analysis. 
Lowe defined four kinds of authentication goals including aliveness, weak agreement, non-injective agreement and injective agreement.

Based on the two threat models from Section~\ref{sec:threatmodel}, the formal code translation generates two Tamarin codes, on which Tamarin Prover will perform formal verification, respectively.
In the neighbor scenario, the attacker sniffs the traffic and gets to know the session key.
Thus, we add a fact to the right-hand side of the session key negotiation rule to indicate that the session key is leaked.
In the tenant scenario, the attacker knows the initial parameters that the client owns in the initial state without sniffing the traffic.
Thus, we add a fact to the right-hand side of the initial rules to indicate that the initial parameters are leaked.

\subsection{Formal Verification}\label{sec:formalverification}
The formal verification module aims to validate the lemmas translated from the security properties on the rules translated from the state machine.
In this module, we apply Tamarin Prover, an off-the-shelf tool for property verification.
However, in the fully automatic mode of Tamarin, not all lemmas can be proved automatically due to the complex state machine, which is a common limitation of Tamarin Prover \cite{tamarinmanual}\cite{basin2018formal}.
This limitation is related to the ranking of unproved goals extracted from the lemma.
To overcome this, Tamarin Prover allows a user to supply heuristics called helping oracle to rank the unproved goals and guide the prove procedure.
Therefore, we design and implement a new ranking strategy on the helping oracle, which is detailed as follows.

The unproved goals extracted from the lemma include validating the source of a state, the existence of an action fact that the attacker knows some parameters (e.g., secret keys, passwords, encrypted parameters), and other goals.
First, we solve the unproved goals to validate the source of a state. Among these goals, the ones that contain a state of a longer trace in the state machine should be solved first, as they can be transformed into the goals that contain a state of a shorter trace.
Second, we solve the unproved goals that validate the existence of an action fact indicating the attacker knows secret key or password.
Third, we solve the unproved goals that validate the existence of an action indicating the attacker knows an encrypted parameter. 
This order can avoid the case of finding no proof path when solving the existence of an action that the attacker knows an encrypted parameter.
Last, we apply the default ranking from Tamarin Prover for the remaining unproved goals.
Our strategy helps Tamarin Prover automatically and efficiently validate the security properties.
For instance, we apply our strategy to prove the authentication lemma of the \texttt{CONNECT} message on the server-side on AWS IoT Core. While the automatic mode will never be terminated, our helping oracle proves that this lemma is false, whose generated proof only costs 13 steps.
As a result, our formal verification module is fully automatic thanks to the proposed helping oracle.
\subsection{Extension for New Types of MPs} \label{sec:extending}

\system has built-in support for security analysis on any customized implementation of MQTT, CoAP and AMQP.
As for new customized MP implementations, the amount of work to be done is to offer three inputs including MP traffic, IoT platform documents and communication configurations, which is simple and requires minimum effort.  
We only need expert involvement when we need to analyze a new MP protocol.
First, the message decoder in the message semantics extraction module and the adapter in the interaction logic extraction module need to be re-implemented according to the types and formats of the messages in the new MP.
\ignore{First, the parameter decoder from message semantics extraction should be modified for extracting the parameters of messages from the MP traffic.
Second, the adapter from interaction logic extraction module should be extended with encoder and decoder for new MP messages.
Thus, the adapter can encode different types of input messages to the broker and decode different response messages from the broker.
The above two modifications requires the knowledge about the types and formats of messages.}
Second, the meta properties for the new MP need to be concluded, which include the necessary messages and parameters that should be authenticated and confidential.
Third, the pre-extracted modeling knowledge from the standard MP specification, e.g., the initial states of the clients and the broker, need to be provided for the formal code translation module.
All the knowledge required is not tied to a specific MP implementation and can be obtained from the public standard specification of the new MP.
Note that the above operations are a one-shot effort for each new MP type.
Actually, in real world, the number of popular MPs is limited and usually stable.
Therefore, \system is directly usable in most scenarios.

\vspace{-3mm}
\section{Evaluations}
\vspace{-2mm}

In this section, we utilize \texttt{MPInspector} to explore ten implementations of MQTT, CoAP and AMQP on nine leading IoT platforms.
We aim to answer the following research questions:

\noindent$\bullet$ \textbf{RQ1:} How well do MP implementations on different platforms follow the security properties? 

\noindent$\bullet$ \textbf{RQ2:}  What are the reasons for property violations?

\noindent$\bullet$ \textbf{RQ3:}  What kind of attacks can be triggered based on property violations?

\noindent$\bullet$ \textbf{RQ4:} How efficient and accurate is \texttt{MPInspector}? 


\subsection{Experiment Settings}
We perform our experiments on a laptop with a 2.6 GHz 2-core Intel i5 processor and 8GB RAM, using Oracle Java Runtime version 1.8 (64 bit) in its default settings.






\noindent\textbf{Evaluation subjects.}
To examine the effectiveness of \system, we evaluate ten MP implementations from nine leading commercial IoT platforms \cite{competitivelandscape}, which are shown in Table~\ref{tab:sarvalidate}.
These implementations cover three main types of MPs, MQTT (including the widely adopted version V3.1.1 and the latest version V5.0), CoAP and AMQP V1.0.
We perform our analysis by buying SaaS applications for MP from the IoT platforms so that the analysis can cover more devices in the real world that use these SaaS applications (see Section~\ref{sec:cloudbasediotplatforms}).

Among the ten evaluated MP implementations, five of them adopt SSL/TLS mechanism, including MQTT on Google IoT Core \cite{googleiotcore}, Azure IoT Hub\cite{azureiothub}, AWS IoT Core\cite{awsiotcore}, Bosch IoT Hub\cite{boschiothub}, and Aliyun Cloud \cite{aliyuncloud}.
We also analyze the secrecy and authenticity properties of MP implementations without SSL/TLS, including MQTT on Tuya Smart \cite{tuya} and Mosquitto \cite{mosquitto}, CoAP on Aliyun Cloud \cite{aliyuncloud} and EMQ X \cite{emqx}, and AMQP on ActiveMQ \cite{activemq}.
Because they are widely adopted by the device manufactures \cite{competitivelandscape} and their security flaws may have a large practical impact.

\noindent\textbf{Validation settings.} We use the client SDK provided by the SaaS application to build potential victims for vulnerabilities and attack validation. As for Tuya Smart, who has acknowledged our findings, we further validate our findings on the real devices under their permission. We also build up scripts based on JavaScript to exploit the vulnerabilities and perform the validation attacks. Performed as an attacker, we manually check those lemmas guided by the attack paths generated by Tamarin Prover. Specifically, we use our scripts to see if we can acknowledge the secret or impersonate the agents in the communication between a server and a client.

\noindent\textbf{Ethical consideration.}
Our study conducts active measurement on the real world MP implementations.
As a result, we take several steps to ensure that our experiments are ethically sound and do not result in the disruption of other users and IoT platforms.
First, we test the SaaS appliactions for MP on our own services bought from the IoT platforms, which does not disrupt other users.
Second, when interacting with the broker on the IoT platforms, our messages are based on the normal traffic produced by us in our own SaaS applications, which does not disrupt the IoT platforms. 
Lastly, we validate our attacks on Tuya Smart with our own devices, which does not influence other devices or the platform.



\subsection{Property Validation} \label{sec:propertyvalidation}
This section answers the questions \textbf{RQ1} and \textbf{RQ2}. 
We show the identified property violations in Table~\ref{tab:sarvalidate}, where we find that all MP implementations encounter various authentication and secrecy property violations, and each MP implementation violates at least 18 properties.

\begin{table}[t]
\caption{ An overview of violated properties (noted as Pr.) in the ten MP implementations. For the checked properties, please refer to Table~\ref{tab:secrecyproperties} and Table~\ref{tab:authenticationproperties} in Appendix \ref{app:properties}. \label{tab:sarvalidate}}
\centering
 \scriptsize
\resizebox{\linewidth}{!}{%
\begin{tabular}{|c|c|c|c|c|c|}
\hline
\multirow{2}{*}{\textbf{Platform}} & \multirow{2}{*}{\textbf{MPs}} & \multicolumn{2}{c|}{\textbf{Secrecy Pr.}} & \multicolumn{2}{c|}{\textbf{Authentication Pr.}} \\ \cline{3-6} 
 &  & \begin{tabular}[c]{@{}c@{}}Neighbor\\ Scenario\end{tabular} & \begin{tabular}[c]{@{}c@{}}Tenant\\ Scenario\end{tabular} & \begin{tabular}[c]{@{}c@{}}Neighbor \\ Scenario\end{tabular} & \begin{tabular}[c]{@{}c@{}}Tenant\\ Scenario\end{tabular} \\ \hline
\begin{tabular}[c]{@{}c@{}}Google IoT\\ Core \end{tabular} & \begin{tabular}[c]{@{}c@{}}MQTT\\ V3.1.1\end{tabular} & \textbf{MS\{1,3-6\}} & MS\{1-6\} & MA\{1-9\} & \begin{tabular}[c]{@{}c@{}}MA\{1,3,\\ 5,7,9\}\end{tabular} \\ \hline
\begin{tabular}[c]{@{}c@{}}AWS IoT\\ Core \end{tabular} & \begin{tabular}[c]{@{}c@{}}MQTT\\ V3.1.1\end{tabular} & MS\{1, 3-6\} & MS\{1, 3-6\} & MA\{1-10\} & \begin{tabular}[c]{@{}c@{}}MA\{1,3,\\ 5,7,9-10\}\end{tabular} \\ \hline
\begin{tabular}[c]{@{}c@{}}Azure IoT\\ Hub \end{tabular} & \begin{tabular}[c]{@{}c@{}}MQTT\\ V3.1.1\end{tabular} & \textbf{MS\{1, 3-6\}} & MS\{1-6\} & MA\{1-9\} & \begin{tabular}[c]{@{}c@{}}MA\{1,3,\\ 5,7,9\}\end{tabular} \\ \hline
\begin{tabular}[c]{@{}c@{}}Bosch IoT\\ Hub \end{tabular} & \begin{tabular}[c]{@{}c@{}}MQTT\\ V3.1.1\end{tabular} & MS\{1, 3-6\} & MS\{1, 3-6\} & MA\{1-9\} & \begin{tabular}[c]{@{}c@{}}MA\{1,3,\\ 5,7,9\}\end{tabular} \\ \hline
\begin{tabular}[c]{@{}c@{}}Aliyun\\ Cloud \end{tabular} & \begin{tabular}[c]{@{}c@{}}MQTT\\ V3.1.1\end{tabular} & \textbf{MS\{1, 3-6\}} & MS\{1-6\} & MA\{1-9\} & \begin{tabular}[c]{@{}c@{}}MA\{1,3,\\ 5,7,9\}\end{tabular} \\ \hline
\begin{tabular}[c]{@{}c@{}}Tuya\\ Smart\end{tabular} & \begin{tabular}[c]{@{}c@{}}MQTT\\ V3.1.1\end{tabular} & \textbf{MS\{1,  3-5\}} & MS\{1-6\} & \textbf{\begin{tabular}[c]{@{}c@{}}MA\{1-6,\\ 8-9\}\end{tabular}} & \begin{tabular}[c]{@{}c@{}}MA\{1,3,\\ 5,7,9\}\end{tabular} \\ \hline
\begin{tabular}[c]{@{}c@{}}Mosquitto\\ \end{tabular} & \begin{tabular}[c]{@{}c@{}}MQTT\\ V5.0\end{tabular} & \begin{tabular}[c]{@{}c@{}}MS\{1,\\ 3-9\}\end{tabular} & MS\{1, 3-9\} & MA\{1-11\} & \begin{tabular}[c]{@{}c@{}}MA\{1,3,\\ 5,7,9-11\}\end{tabular} \\ \hline
\begin{tabular}[c]{@{}c@{}}EMQ X\\ \end{tabular} & CoAP & CS\{1-6\} & CS\{1-6\} & CA\{1-4\} & CA\{1,3\} \\ \hline
\begin{tabular}[c]{@{}c@{}}Aliyun\\ Cloud \end{tabular} & CoAP & \textbf{\begin{tabular}[c]{@{}c@{}}CS\{1-4,\\ 7, 9-10\}\end{tabular}} & \begin{tabular}[c]{@{}c@{}}CS\{1-4,\\ 7-10\}\end{tabular} & \textbf{CA\{5-6, 8\}} & CA\{5,7\} \\ \hline
\begin{tabular}[c]{@{}c@{}}ActiveMQ\\ \end{tabular} & \begin{tabular}[c]{@{}c@{}}AMQP\\ V1.0\end{tabular} & AS\{1-5\} & AS\{1-5\} & AA\{1-13\} & \begin{tabular}[c]{@{}c@{}}AA\{1,3,5,\\ 7, 9,11,13\}\end{tabular} \\ \hline
\end{tabular}
}
\end{table}

\subsubsection{Neighbor Scenario}

In the neighbor scenario, \texttt{MPInspector} identifies that three out of the ten MP implementations (Mosquitto, EMQ X, and ActiveMQ) violate all the security properties. 
The rest of these implementations violate at least ten secrecy properties and five authentication properties.  

\noindent\textbf{Secrecy properties.}
We identify that five MP implementations (MQTT on Tuya Smart and Mosquitto, CoAP on Aliyun Cloud and EMQ X, AMQP on ActiveMQ) support transmitting messages in plain text. The other five MP implementations (MQTT on Google IoT Core, AWS IoT Core, Azure IoT Hub, Bosch IoT Hub, and Aliyun Cloud) 
adopt SSL/TLS but are still facing SSL/TLS interception risks because of wrong configurations. In addition, their messages can still be decrypted by man-in-the-middle attacks.
As a result, for all the ten implementations, \system identifies that the secrecy properties for the parameters without additional encryption are all failed.
Below we discuss the secrecy properties on the parameters with additional encryption.
Five MP implementations (Google IoT Core, Azure IoT Hub, Bosch IoT Hub, Aliyun Cloud, and Tuya Smart) deploy additional encryption on some of their parameters.
Among them, Google IoT Core and Azure IoT Hub use a secret key to generate JWT and SAS tokens, which are valid before the expired time. In the neighbor scenario, the unexpired token can be reused by an attacker.
Aliyun Cloud encrypts a client's secrets with timestamps by a secret key. The CoAP implementation in Aliyun Cloud additionally encrypts the payload in the \texttt{POST\_PUBLISH} message with a timestamp by a secret key.
However, \system validates that the timestamp is not checked by the server, which suggests that the password and payload in Aliyun Cloud can be reused as well.
Tuya Smart uses a secret key to encrypt a client's password in the \texttt{CONNECT} message and encrypt the payload with a timestamp in the \texttt{PUBLISH} message.
\system identifies that Tuya Smart satisfies the secrecy property for \texttt{PUBLISH Payload} but fails the secrecy for the password.

\noindent\textbf{Authentication properties.} 
\system validates authentication properties on both the client side and the server side.
Table~\ref{tab:sarvalidate} shows the overview of the authentication property violations detected by \texttt{MPInspector}.

From the results, three MP implementations without any authentication mechanism (Mosquitto, EMQ X, and ActiveMQ) fail the aliveness goals of all authentication lemmas.
Five MP implementations including Google IoT Core, AWS IoT Core, Azure IoT Hub, Bosch IoT Hub, and Aliyun Cloud that adopt SSL/TLS satisfy the non-injective goals on the \texttt{CONNECT} message of the server side.
However, they still fail the non-injective goals on the \texttt{CONNECT} message because of SSL/TLS interception.
Their other messages (\texttt{SUBSCRIBE}, \texttt{UNSUBSCRIBE}, \texttt{PUBLISH}, \texttt{DISCONNECT} messages) without authentication fail the aliveness goals.
The rest two implementations (MQTT on Tuya Smart and CoAP on Aliyun Cloud) do not adopt SSL/TLS but adopt an encryption mechanism on their messages. 
For Tuya Smart, the \texttt{CONNECT} message on the server side satisfies the aliveness goal but fails the weak agreement goal.
Therefore, even though the password is encrypted by a secret key, the attacker can still sniff and reuse on the \texttt{CONNECT} message.
For Aliyun Cloud's CoAP implementation, it has encryption but does not check the timestamp in \texttt{CON\_POSTAUTH} and \texttt{CON\_POSTPUBLISH} messages.
Therefore, an attacker can connect with the server by replaying the messages he collected from the client previously.
As a result, in Aliyun Cloud, authentications on \texttt{CON\_POSTAUTH} and \texttt{CON\_POSTPUBLISH} messages satisfy the weak agreement goal but fail the non-injective goal.

\subsubsection{Tenant Scenario}
In the tenant scenario, \system has identified that all the secrecy properties are violated in all the ten implementations.
The reason is that the attacker can impersonate the victim to connect with the server and accept all the messages from the server. 
For authentication properties, \texttt{MPInspector} identifies that all the ten implementations violate all the properties on the server side, but meet the properties on the device side.
This is due to the differences of the attacker's capabilities to control the device side and the server side.
On the device side, the attacker cannot steal the session key as he may not be in the same network with the victim.
While on the server side, the attacker can create a fake client to connect with the server using the identities he created when he has access to the device.
Then, the server would recognize the fake client as a legitimate one, which allows the attacker to break all the authentication goals. 



\vspace{-3mm}
\subsection{Attacks based on the Property Violations}\label{sec:findings}
This section answers the question \textbf{RQ3}.
Based on the property violations, we uncover eleven kinds of attacks on the ten MP implementations and display the overview in Table~\ref{findings}.
We find that the examined MP implementations are all vulnerable under the two attack scenarios.
Each platform is vulnerable to at least one attack, and on average 2.8 attacks.
These attacks have serious consequences, such as sensitive data leakage and malicious message injection. 
We introduce six attacks below (more attacks are available in \cite{mpinspector}).

\begin{table}[t]
\centering
 \caption{ Attacks and relevant property (noted as Related Pr.) violations (\protect\includegraphics[scale=0.015]{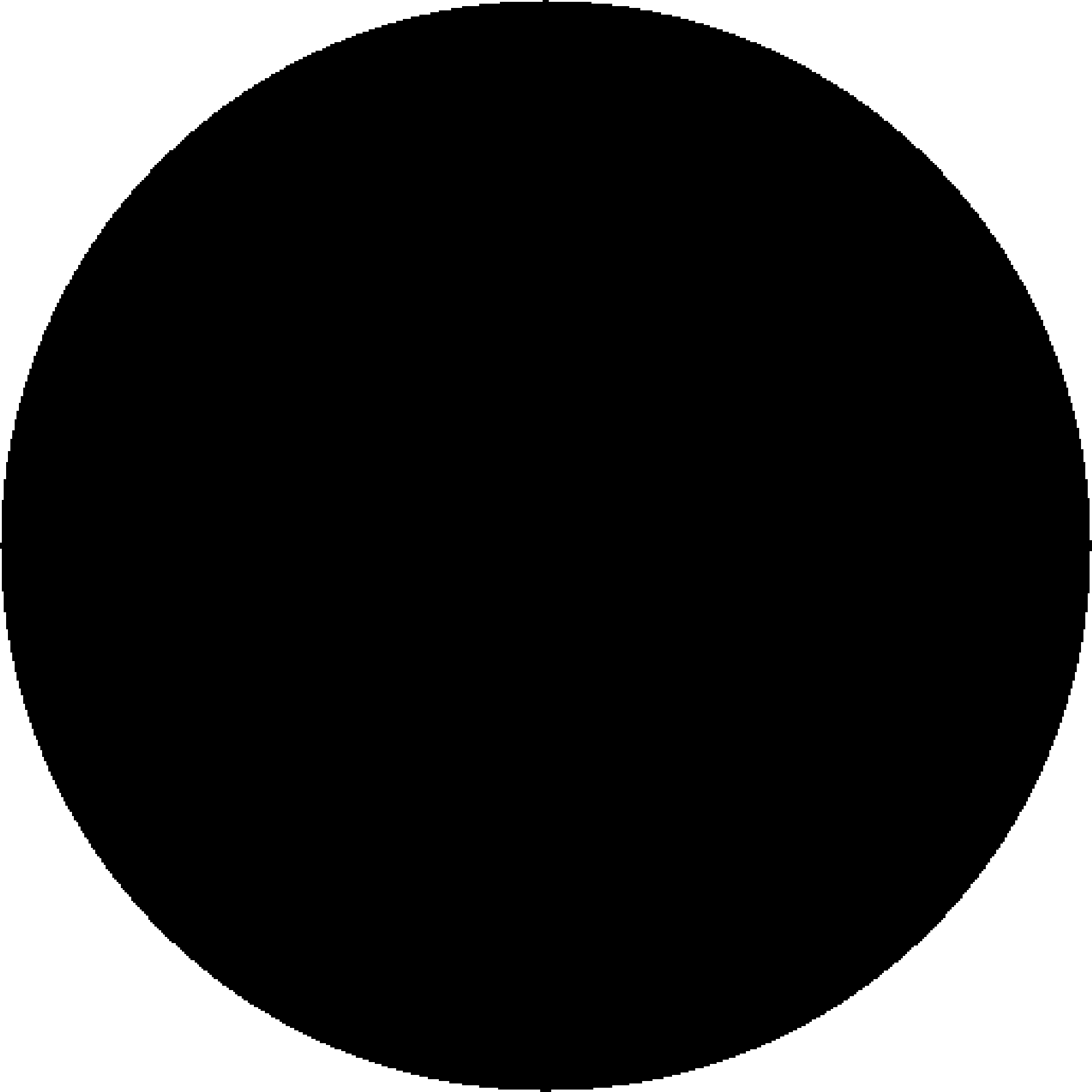}=validated, \protect\includegraphics[scale=0.015]{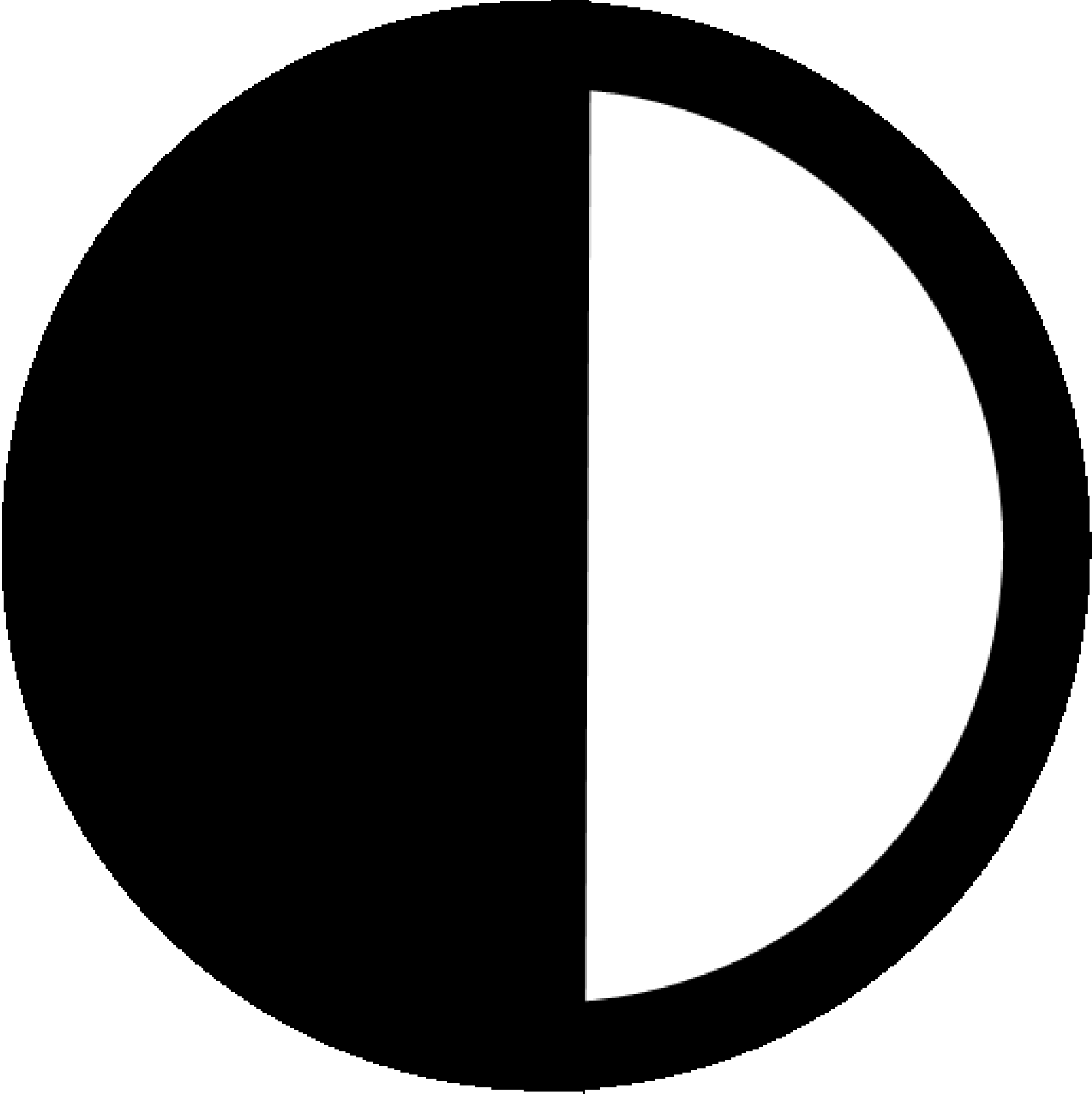}=partially validated).}
 \label{findings}
  \scriptsize
\resizebox{\linewidth}{!}{%
\begin{tabular}{|c|c|c|c|c|}
\hline
\textbf{\begin{tabular}[c]{@{}c@{}}Neighbor\\ Scenario\end{tabular}}  & \textbf{\begin{tabular}[c]{@{}c@{}}Affected\\ Protocol\end{tabular}} & \textbf{\begin{tabular}[c]{@{}c@{}}Affected\\ Platforms\end{tabular}} & \textbf{Related Pr.} & \textbf{\begin{tabular}[c]{@{}c@{}}Verified\end{tabular}} \\ \hline
 &  &  &  &  \\
 &  &  &  &  \\
\multirow{-3}{*}{\begin{tabular}[c]{@{}c@{}}Man-in-\\ the-middle\end{tabular}} & \multirow{-3}{*}{All protocols} & \multirow{-3}{*}{All platforms} & \multirow{-3}{*}{\begin{tabular}[c]{@{}c@{}}MA\{1-9\},\\ AA\{1-13\}, \\ CA\{1-8\}\end{tabular}} & \multirow{-3}{*}{\includegraphics[scale=0.015]{pic/mark_all-eps-converted-to.pdf}} \\ \hline
 &  & AWS IoT Core &  &  \\ \cline{3-3}
 & \multirow{-2}{*}{MQTT V3.1.1} & Tuya Smart & \multirow{-2}{*}{MA\{1-9\}} & \multirow{-2}{*}{\includegraphics[scale=0.015]{pic/mark_all-eps-converted-to.pdf}} \\ \cline{2-5} 
 &  &  &  &  \\
 & \multirow{-2}{*}{MQTT V5.0} & \multirow{-2}{*}{Mosquirro} & \multirow{-2}{*}{\begin{tabular}[c]{@{}c@{}}MA\{1-9\},  \\ MA\{10-11\}\end{tabular}} & \multirow{-2}{*}{\includegraphics[scale=0.015]{pic/mark_all-eps-converted-to.pdf}} \\ \cline{2-5} 
 &  &  &  &  \\
 & \multirow{-2}{*}{CoAP} & \multirow{-2}{*}{EMQ X} & \multirow{-2}{*}{CA\{1-4\}} & \multirow{-2}{*}{\includegraphics[scale=0.015]{pic/mark_all-eps-converted-to.pdf}} \\ \cline{2-5} 
 &  &  &  &  \\
\multirow{-8}{*}{\begin{tabular}[c]{@{}c@{}}Replay\\ Attack\end{tabular}} & \multirow{-2}{*}{AMQP V1.0} & \multirow{-2}{*}{ActiveMQ} & \multirow{-2}{*}{AA\{1-13\}} & \multirow{-2}{*}{\includegraphics[scale=0.015]{pic/mark_all-eps-converted-to.pdf}} \\ \hline
 &  &  &  &  \\
\multirow{-2}{*}{\begin{tabular}[c]{@{}c@{}}Transfer Sync.\\ Failure\end{tabular}} & \multirow{-2}{*}{AMQP V1.0} & \multirow{-2}{*}{ActiveMQ} & \multirow{-2}{*}{AA\{1-9\}} & \multirow{-2}{*}{\includegraphics[scale=0.015]{pic/mark_all-eps-converted-to.pdf}} \\ \hline
\textbf{\begin{tabular}[c]{@{}c@{}}Tenant\\ Scenario\end{tabular}} & \textbf{\begin{tabular}[c]{@{}c@{}}Affected\\ Protocol\end{tabular}} & \textbf{\begin{tabular}[c]{@{}c@{}}Affected\\ Platforms\end{tabular}} & \textbf{Related Pr.} & \textbf{Verified} \\ \hline
 &  & Google IoT Core &  &  \\ \cline{3-3}
 &  & Azure IoT Hub &  &  \\ \cline{3-3}
 &  & AWS IoT Core &  &  \\ \cline{3-3}
 & \multirow{-4}{*}{MQTT V3.1.1} & Aliyun Cloud &  &  \\ \cline{2-3}
 & MQTT V5.0 & Mosquitto & \multirow{-5}{*}{\begin{tabular}[c]{@{}c@{}}MS\{1-7\},\\ MA\{1,3,5,\\ 7,9\}\end{tabular}} & \multirow{-5}{*}{\includegraphics[scale=0.015]{pic/mark_all-eps-converted-to.pdf}} \\ \cline{2-5} 
 &  &  &  &  \\
 &  &  &  &  \\
 & \multirow{-3}{*}{AMQP V1.0} & \multirow{-3}{*}{ActiveMQ} & \multirow{-3}{*}{\begin{tabular}[c]{@{}c@{}}AS\{1-5\}, \\ AA\{1.3,5,\\ 7,9,11,13\}\end{tabular}} & \multirow{-3}{*}{\includegraphics[scale=0.015]{pic/mark_all-eps-converted-to.pdf}} \\ \cline{2-5} 
 &  & EMQ X &  &  \\ \cline{3-3}
\multirow{-10}{*}{\begin{tabular}[c]{@{}c@{}}Client\\ Identity\\ Hijacking\end{tabular}} & \multirow{-2}{*}{CoAP} & Aliyun Cloud & \multirow{-2}{*}{\begin{tabular}[c]{@{}c@{}}CS\{1-11\},\\ CA\{1,3,5,7\}\end{tabular}} & \multirow{-2}{*}{\includegraphics[scale=0.015]{pic/mark_all-eps-converted-to.pdf}} \\ \hline
 &  & EMQ X &  &  \\ \cline{3-3}
\multirow{-2}{*}{\begin{tabular}[c]{@{}c@{}}Reflection\\ Attack\end{tabular}} & \multirow{-2}{*}{CoAP} & Aliyun Cloud & \multirow{-2}{*}{\begin{tabular}[c]{@{}c@{}}CS1,\\ CA\{1,3,5,7\}\end{tabular}} & \multirow{-2}{*}{\includegraphics[scale=0.015]{pic/mark_half-eps-converted-to.pdf}} \\ \hline
 &  &  &  &  \\
 & \multirow{-2}{*}{MQTT V3.1.1} & \multirow{-2}{*}{AWS IoT Core} & \multirow{-2}{*}{\begin{tabular}[c]{@{}c@{}}MS\{5,7-9\},\\ MA3\end{tabular}} & \multirow{-2}{*}{\includegraphics[scale=0.015]{pic/mark_all-eps-converted-to.pdf}} \\ \cline{2-5} 
 &  &  &  &  \\
\multirow{-4}{*}{\begin{tabular}[c]{@{}c@{}}Malicious\\ Topic\\ Subscription\end{tabular}} & \multirow{-2}{*}{AMQP V1.0} & \multirow{-2}{*}{ActiveMQ} & \multirow{-2}{*}{\begin{tabular}[c]{@{}c@{}}AS\{2,4\},\\ AA9\end{tabular}} & \multirow{-2}{*}{\includegraphics[scale=0.015]{pic/mark_all-eps-converted-to.pdf}} \\ \hline
 &  &  &  &  \\
 & \multirow{-2}{*}{MQTT V3.1.1} & \multirow{-2}{*}{AWS IoT Core} & \multirow{-2}{*}{\begin{tabular}[c]{@{}c@{}}MS\{5,7-9\},\\ MA7\end{tabular}} & \multirow{-2}{*}{\includegraphics[scale=0.015]{pic/mark_all-eps-converted-to.pdf}} \\ \cline{2-5} 
 &  &  &  &  \\
\multirow{-4}{*}{\begin{tabular}[c]{@{}c@{}}Malicious\\ Topic\\ Publish\end{tabular}} & \multirow{-2}{*}{CoAP} & \multirow{-2}{*}{EMQ X} & \multirow{-2}{*}{CS1,CA3} & \multirow{-2}{*}{\includegraphics[scale=0.015]{pic/mark_all-eps-converted-to.pdf}} \\ \hline
 &  &  &  &  \\
 &  &  &  &  \\
\multirow{-3}{*}{\begin{tabular}[c]{@{}c@{}}Malicious\\ Response\\ Topic Publish\end{tabular}} & \multirow{-3}{*}{MQTT V5.0} & \multirow{-3}{*}{Mosquitto} & \multirow{-3}{*}{\begin{tabular}[c]{@{}c@{}}MS\{5,7-9\},\\ MA7\end{tabular}} & \multirow{-3}{*}{\includegraphics[scale=0.015]{pic/mark_half-eps-converted-to.pdf}} \\ \hline
 &  &  &  &  \\
 & \multirow{-2}{*}{MQTT V3.1.1} & \multirow{-2}{*}{AWS IoT Core} &  &  \\ \cline{2-3}
 &  &  &  &  \\
\multirow{-4}{*}{\begin{tabular}[c]{@{}c@{}}Unauthorized\\ Will\\ Message\end{tabular}} & \multirow{-2}{*}{MQTT V5.0} & \multirow{-2}{*}{Mosquitto} & \multirow{-4}{*}{MA\{1, 10\}} & \multirow{-4}{*}{\includegraphics[scale=0.015]{pic/mark_all-eps-converted-to.pdf}} \\ \hline
 &  &  &  &  \\
 &  &  &  &  \\
\multirow{-3}{*}{\begin{tabular}[c]{@{}c@{}}Unauthorized\\ Retained\\ Message\end{tabular}} & \multirow{-3}{*}{MQTT V5.0} & \multirow{-3}{*}{Mosquitto} & \multirow{-3}{*}{MA\{8, 11\}} & \multirow{-3}{*}{\includegraphics[scale=0.015]{pic/mark_all-eps-converted-to.pdf}} \\ \hline
 &  &  &  &  \\
\multirow{-2}{*}{\begin{tabular}[c]{@{}c@{}}Illegal\\ Occupation\end{tabular}} & \multirow{-2}{*}{AMQP V1.0} & \multirow{-2}{*}{ActiveMQ} & \multirow{-2}{*}{\begin{tabular}[c]{@{}c@{}}AS1, \\ AA\{1, 3\}\end{tabular}} & \multirow{-2}{*}{\includegraphics[scale=0.015]{pic/mark_all-eps-converted-to.pdf}} \\ \hline
\end{tabular}
}
\end{table}

 \subsubsection{Neighbor Scenario Attacks}

\noindent\textbf{Replay attack.} This attack is due to the authentication property violations, which suggests that the server accepts the messages that the client has sent before.
An attacker only needs to collect them and replays them to the server.
We identify that CoAP on EMQ X, AMQP on ActiveMQ, and MQTT on Tuya Smart, AWS IoT Core and Mosquitto are vulnerable to this attack.
We launch this attack on Tuya Smart and Mosquitto by sniffing and collecting the traffic in the local network and replaying them to the server.
As a result, We successfully replay all the messages, including sending commands and telemetry data.

\noindent\textbf{AMQP sync. failure.} \texttt{MPInspector} finds that the client and server in AMQP strictly maintain the message ID called Delivery ID when sending TRANSFER messages.
Utilizing the authentication property violations on AMQP messages, an attacker can kick the victim offline by sending the messages in wrong orders or forging the TRANSFER messages with synchronized Delivery ID using the victim's identity.
We identify the attack on ActiveMQ. We develop an attack script using Ettercap and have successfully launched the attack on ActiveMQ. 

\subsubsection{Tenant Scenario Attacks}

\noindent\textbf{Client identity hijacking.}
\system detects that the secrecy properties on the device side are all violated in the tenant scenario.
Additionally, the server side authentication properties are also violated. 
This suggests that an attacker can impersonate the victim device using its identity to connect to the server.
We name this attack as client identity hijacking. Especially, \system detects that the MQTT implementations disconnect the existing client when the server receives a second connection request with the same ClientID.
Therefore, an attacker can use the victim's identity to connect to the server and kick the victim offline.
Last, an attacker can impersonate the device to send messages to the server.
We successfully launch this attack on Google IoT Core, AWS IoT Core, Aliyun Cloud, Tuya Smart, Mosquitto, and ActiveMQ.
Additionally, we find that the attacker once obtains the credentials of the client, he can perform this attack for a long time as these IoT platforms hard-code the credentials of the clients into device SDKs and cannot dynamically revoke or grant new credentials.

\begin{figure}[htbp]
    \centering
    \includegraphics[width=0.28\textwidth]{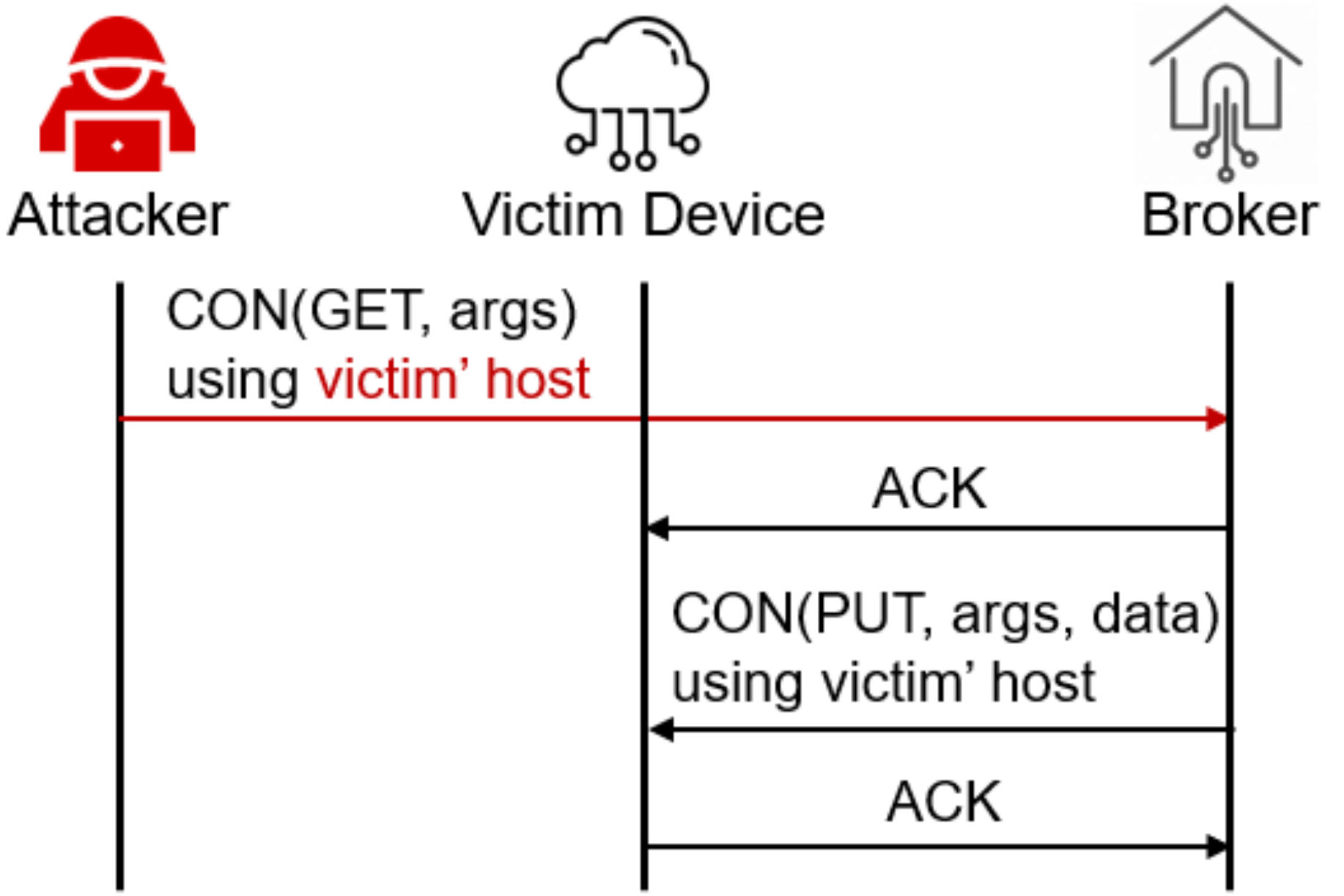}
    \vspace{-4mm}
    \caption{CoAP reflection attack.}
    \label{fig:coapreflection}
\end{figure}

\noindent\textbf{Reflection attack.} 
The reflection attack is specific to the CoAP protocol, which is running over UDP.
Utilizing the secrecy property and authentication property violations on the MP implementations in the server side, an attacker can forge messages using the victim's IP address to send to the server.
The workflow is shown in Figure~\ref{fig:coapreflection}.
We identify the attack of CoAP on Aliyun Cloud and EMQ X. As a consequence, an attacker can forge a fake state to deceive the server.
Also, the attacker can forge a message to get a considerable amount of messages sent to the victim and cause a DoS.
To validate the attack, we use source address spoofing to forge a CoAP message, and the victim successfully receives the unexpected response message.
According to our experiments, the amplification reflection rates are 2.25 in Aliyun Cloud and 0.68 in EMQ X, respectively.
The amplification reflection rate here is a conservative estimation because we adopt the basic configuration, where the broker only returns the response code without the device data.

\begin{figure}[htbp]
    \centering
    \includegraphics[width=0.4\textwidth]{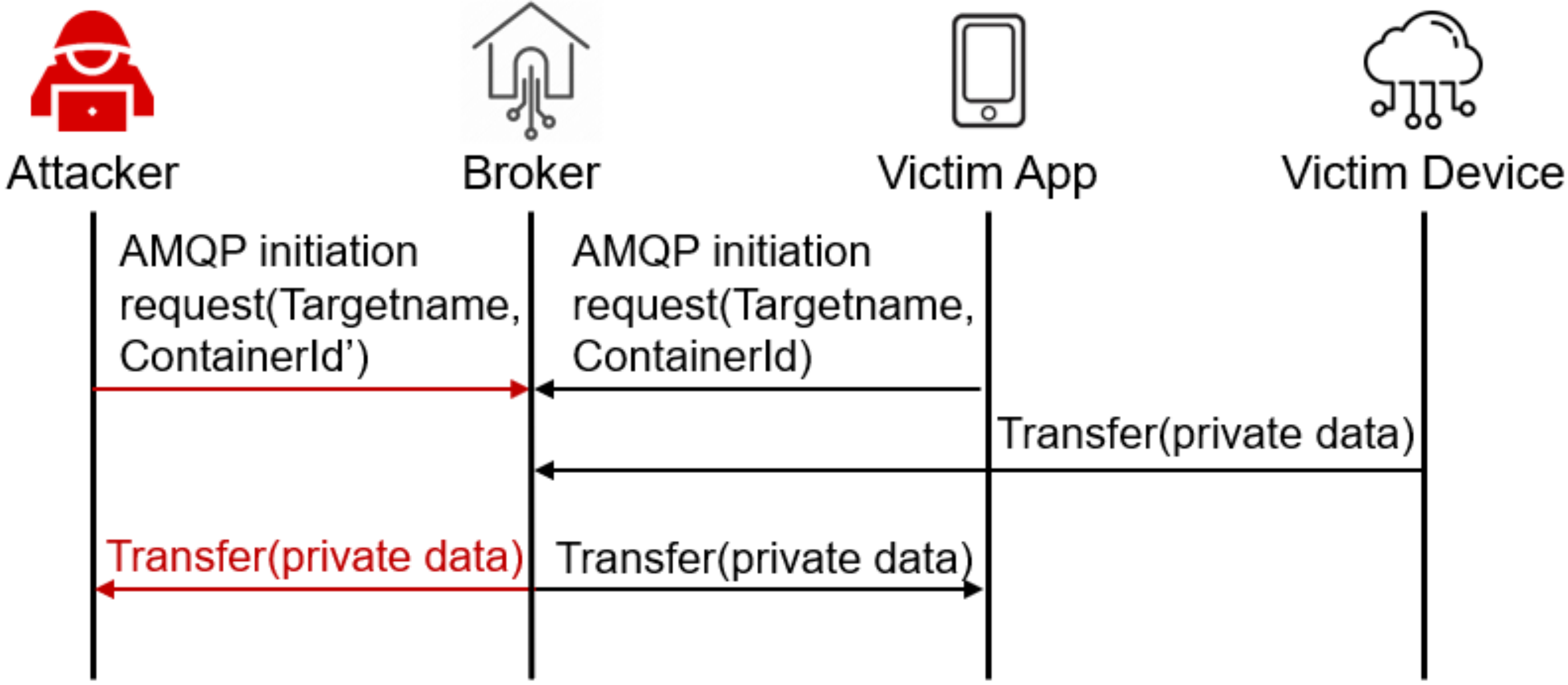}
   \vspace{-4mm}
    \caption{AMQP malicious topic subscription on ActiveMQ.}
    \label{fig:amqpsecretdataleak}
\end{figure}

\noindent\textbf{Malicious topic subscription.}
Because of the secrecy property violation on the topic name and the authentication property violation in the \texttt{SUBSCRIBE} message, an attacker can subscribe to the victim's topic using his own identity.
Taking AMQP as an example, as shown in Figure~\ref{fig:amqpsecretdataleak}, an attacker uses his own identity \texttt{ContainerId} to subscribe to the victim's topic, which is denoted as the target node.
When the victim device sends its secret data, the broker transfers the secret data to the attacker.
We identify this attack on AWS IoT Core and ActiveMQ and further validate this attack successfully. 

\begin{figure}[htbp]
    \centering
    \includegraphics[width=0.4\textwidth]{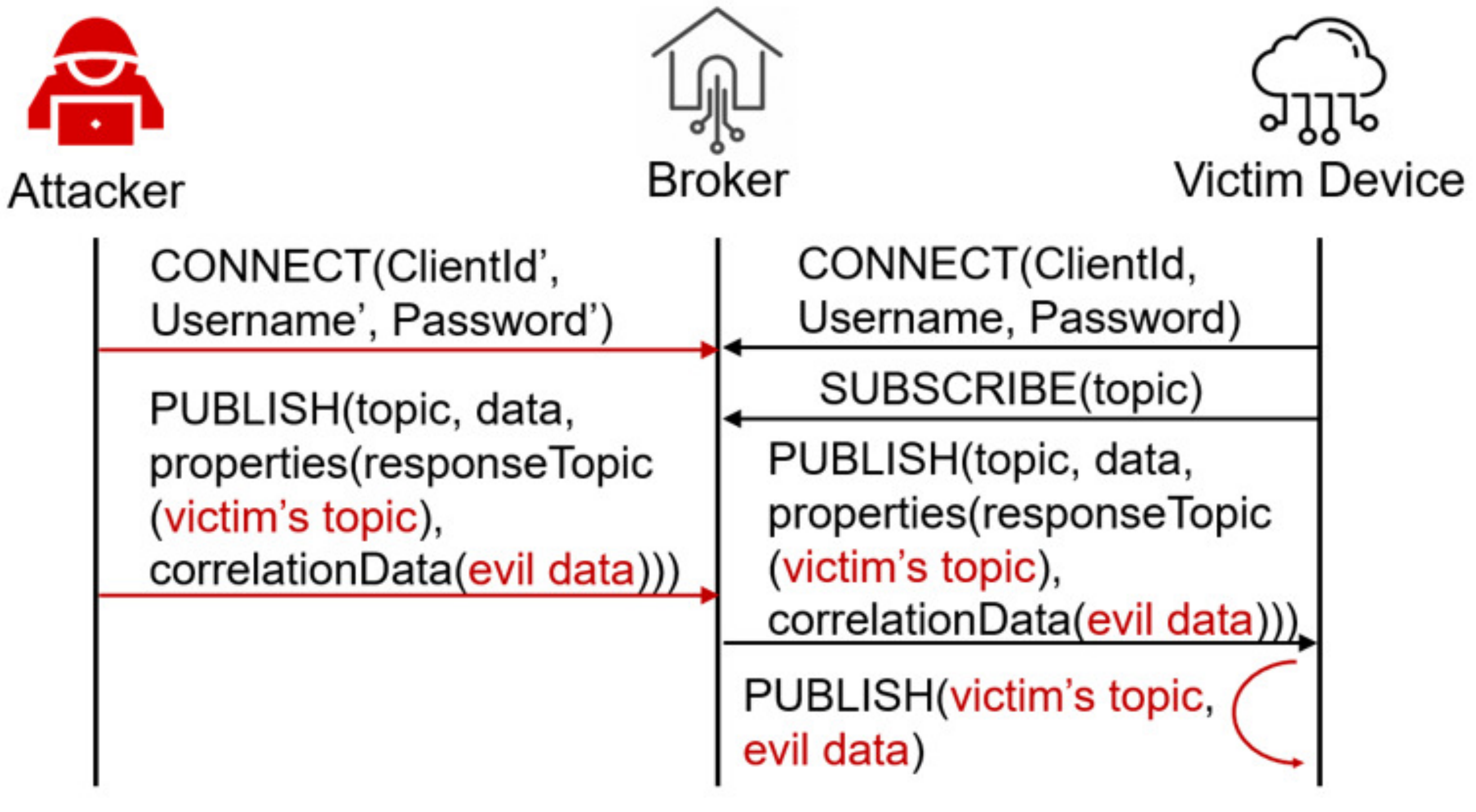}
    \vspace{-4mm}
    \caption{MQTT V5.0 unauthorized response topic publish.}
    \label{fig:mqtt5attacks}
\end{figure}

\noindent\textbf{Unauthorized response message. }
This attack works for the new request/response mechanism introduced by MQTT V5.0.
This mechanism allows the client to publish a message with a response topic and the correlation data.
The client who receives this message publishes the correlation data to the response topic.
However, an attacker can publish with an unauthorized response topic to the victim, as shown in Figure~\ref{fig:mqtt5attacks}.
This attack is based on the secrecy property violation on the victim's topic. It is identified on Mosquitto as it supports MQTT V5.0.
To validate the attack, we use our script to simulate the victim and accomplish the request/response mechanism.
We successfully launch the attack as the broker does not check the authenticity of the response topic.

\begin{figure}[htbp]
    \centering
    \includegraphics[width=0.38\textwidth]{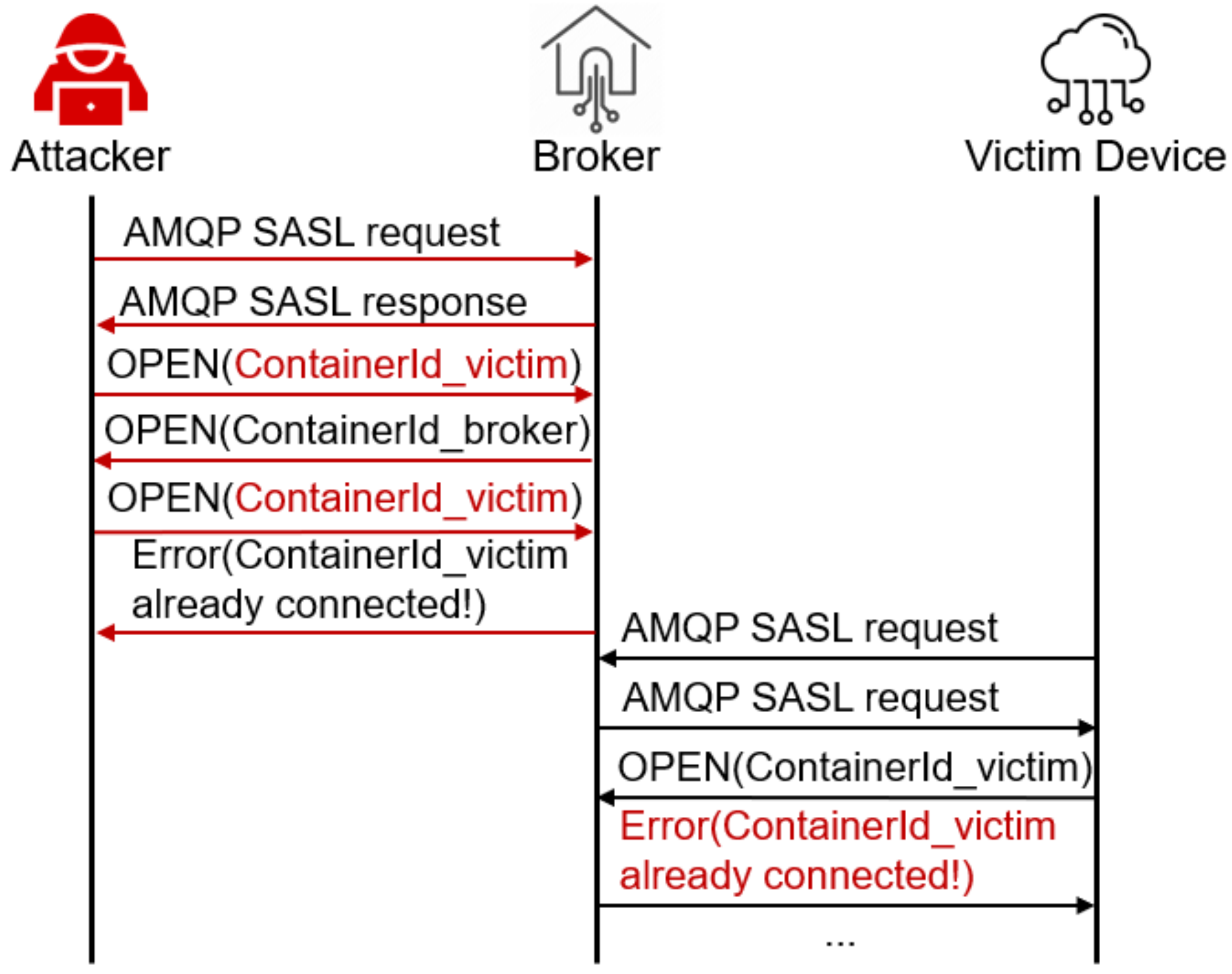}
    \vspace{-4mm}
    \caption{AMQP illegal occupation.}
    \label{fig:amqpio}
\end{figure}

\noindent\textbf{Illegal occupation.} An attacker can exploit the violated secrecy property on the victim's \texttt{ContainerId} and the violated authentication property to perform illegal occupation attacks on AMQP.
The server that receives duplicate \texttt{OPEN} messages with the same \texttt{ContainerId} of the victim closes the connection without updating the session state.
When the client reconnects to the server, the server believes that the client with \texttt{ContainerId} is online and rejects the victim's connection request.
We identify this attack on ActiveMQ, and we believe this attack is severe.
As shown in Figure~\ref{fig:amqpio}, an attacker can collect victims's \texttt{ContainerIds} to perform this attack, and make plenty of victims out of service unless the broker resets.
We use our script to launch this attack successfully, and the target victim cannot connect to the broker anymore.

\vspace{-3mm}
\subsubsection{Comparisons with Burglars' IoT Paradise Paper}\label{sec:comparison}
In \cite{jia2020burglars}, Jia \textit{et al.} performed a manual analysis on MQTT manually.
We compare \system with \cite{jia2020burglars}, which is shown in Table~\ref{tab:findingcomparison}.
Our framework is automatic while \cite{jia2020burglars} only analyzed MQTT manually.
In addition, \system covers four prominent MPs including MQTT V3.1.1, MQTT V5.0, CoAP and AMQP V1.0 while \cite{jia2020burglars} only analyzed MQTT V3.1.1.
As for MQTT V3.1.1, we find four new attacks that \cite{jia2020burglars} did not cover. We consider the neighbor scenario and the tenant scenario and \cite{jia2020burglars} only considered the latter.
There are two attacks in \cite{jia2020burglars} that \system does not cover.
However, these two attacks are either not MQTT's implementation flaw or related to the understanding of bit wise parameters, which are out of the current design focus of \system.
Instead, \system is mainly designed for logic flaw analysis on MP implementations.
In conclusion, compared with the previous work \cite{jia2020burglars}, \system is an automatic approach, covers more MPs and reveals four more new attacks.

\begin{table}[htb]
\centering
 \caption{ Our results compared with \cite{jia2020burglars} in MQTT ((\protect\includegraphics[scale=0.015]{pic/mark_all-eps-converted-to.pdf}=detected, \protect\includegraphics[scale=0.014]{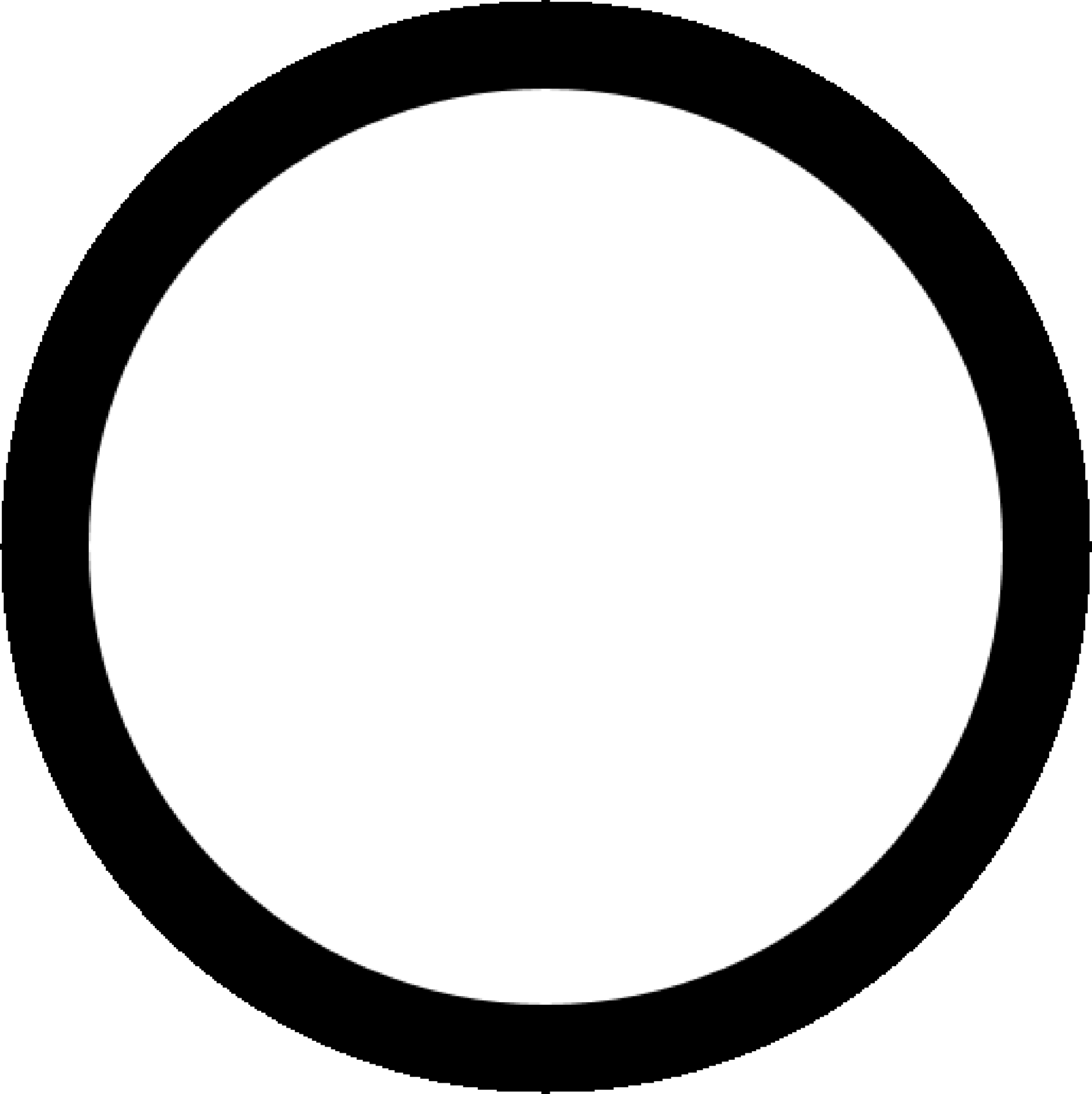}=not detected)). }
 \textsubscript{}
 \label{tab:findingcomparison}
 \scriptsize
 \begin{tabular}{|c|c|c|c|}
\hline
\textbf{Scenario} & \textbf{Types of Attacks} & \textbf{\begin{tabular}[c]{@{}c@{}} \cite{jia2020burglars}\end{tabular}} & \textbf{MPInspector} \\ \hline
\multirow{2}{*}{\begin{tabular}[c]{@{}c@{}}Neighbor\\ Scenario\end{tabular}} & \textbf{Man in the Middle} & \includegraphics[width=0.015\textwidth]{pic/mark_null-eps-converted-to.pdf} & \includegraphics[scale=0.014]{pic/mark_all-eps-converted-to.pdf} \\ \cline{2-4} 
 & \textbf{Replay Attack} & \includegraphics[scale=0.014]{pic/mark_null-eps-converted-to.pdf} & \includegraphics[scale=0.014]{pic/mark_all-eps-converted-to.pdf} \\ \hline
\multirow{9}{*}{\begin{tabular}[c]{@{}c@{}}Tenant\\ Scenario\end{tabular}} & \textbf{Unauthorized Will Message} & \includegraphics[width=0.014\textwidth]{pic/mark_all-eps-converted-to.pdf} & \includegraphics[scale=0.014]{pic/mark_all-eps-converted-to.pdf} \\ \cline{2-4} 
 & \textbf{Malicious Retained Message} & \includegraphics[scale=0.014]{pic/mark_all-eps-converted-to.pdf} & \includegraphics[scale=0.014]{pic/mark_all-eps-converted-to.pdf} \\ \cline{2-4} 
 & \textbf{Client Identity Hijacking} & \includegraphics[scale=0.014]{pic/mark_all-eps-converted-to.pdf} & \includegraphics[scale=0.014]{pic/mark_all-eps-converted-to.pdf} \\ \cline{2-4} 
 & \textbf{ClientID identification} & \includegraphics[scale=0.014]{pic/mark_all-eps-converted-to.pdf} & \includegraphics[scale=0.014]{pic/mark_null-eps-converted-to.pdf} \\ \cline{2-4} 
 & \textbf{Malicious Topic Subscription} & \includegraphics[scale=0.014]{pic/mark_all-eps-converted-to.pdf} & \includegraphics[scale=0.014]{pic/mark_all-eps-converted-to.pdf} \\ \cline{2-4} 
 & \textbf{Malicious Topic Publish} & \includegraphics[scale=0.015]{pic/mark_null-eps-converted-to.pdf} & \includegraphics[scale=0.015]{pic/mark_all-eps-converted-to.pdf} \\ \cline{2-4} 
 & \textbf{Wildcard-topic Subscription} & \includegraphics[scale=0.015]{pic/mark_all-eps-converted-to.pdf} & \includegraphics[scale=0.015]{pic/mark_null-eps-converted-to.pdf} \\ \cline{2-4} 
 & \textbf{\begin{tabular}[c]{@{}c@{}}Unauthorized Response\\ Topic Publish\end{tabular}} & \includegraphics[scale=0.015]{pic/mark_null-eps-converted-to.pdf} & \includegraphics[scale=0.015]{pic/mark_all-eps-converted-to.pdf} \\ \hline
 \end{tabular}
\end{table}

 \vspace{-4mm}
\subsection{Performance}
This section answers question \textbf{RQ4}. We evaluate the performance of \texttt{MPInspector} from three perspectives: (1) state machine modeling, (2) property violation detection, and (3) performance overhead.

\noindent\textbf{Evaluation on state machine modeling.} 
The state machine modeling includes message semantics extraction, interaction logic extraction and formal code translation.
We first evaluate the performance of \system on message semantics extraction on the ten tested MP implementations.
As MP implementations are closed-sourced, it is difficult to get the ground truth of the message semantics for the real MP implementations. Thus, we invite 45 experts with abundant protocol and software reverse engineering experiences to manually validate our results.
Since recovering the full message semantics depends on the amount of collected MP traffic and the quality of IoT platform documents, the experts are instructed to only focus on checking the correctness of each parameter semantics extracted by \system by checking all the available traffic and documents.
Thus, as a precaution, we only report the precision.
As a result, the precision of message semantics extraction on Aliyun Cloud is 96\%, while the precision on other IoT platforms is all 100\%.
As the value of parameter \texttt{ClientID} from Aliyun Cloud includes some irregular characters, our method cannot handle them and mistakenly extracts wrong terms of the parameter.
Additionally, to prove the effectiveness of our NLP-based semantics extraction, we further collect the documents from 20 popular IoT platforms \cite{iotplatformlandscape} for evaluation.
Similarly, our invited experts manually verify the correctness of each extracted parameter semantics by examining the collected documents.
Our method yields 94.87\% precision.  
Our method fails to extract the semantics of some parameters, because the sentences that contain these parameter semantics do not belong to the considered sentence types in Section~\ref{sec:parametersemanticsextraction} and they need to be extracted from several sentences.
For more details on precision of message semantics extraction, please refer to Table~\ref{tab:performance}. 

As for interaction logic extraction, we choose four MP implementations for the evaluation, including Mosquitto, EMQ X, ActiveMQ, and Tuya Smart. The first three are chosen because they are open-source, thus our experts can refer to their code for the ground truth.
Although Tuya Smart is not open-source, with the help of their security team, we can manually review and confirm the result of Tuya Smart.
We cannot validate the other six platforms as we do not have access to their source code.
The validation shows that the state machines learned by \system are consistent with these four implementations. 
As for model translation, we successfully translate all MP state machines into Tamarin code and validate that the codes can successfully run.

\begin{table*}[htb]
\caption{Performance overhead of \system. 
\label{tab:performance}}
\centering
\scriptsize 
\begin{tabular}{|c|c|c|c|c|c|c|c|c|c|c|c|}
\hline
\multirow{2}{*}{\textbf{IoT Platform}} & \multirow{2}{*}{\textbf{MP}} & \multicolumn{2}{c|}{\textbf{\begin{tabular}[c]{@{}c@{}}Message semantics \\ Extraction\end{tabular}}} & \multicolumn{6}{c|}{\textbf{\begin{tabular}[c]{@{}c@{}}Interaction Logic\\ Extraction\end{tabular}}} & \textbf{\begin{tabular}[c]{@{}c@{}}Formal code \\ Translation\end{tabular}} & \multirow{2}{*}{\textbf{\begin{tabular}[c]{@{}c@{}}Total\\ Time\\ (h:mm)\end{tabular}}} \\ \cline{3-11}
 &  & Time (ms) & Precision & States & Time Delay & \begin{tabular}[c]{@{}c@{}}\# of Input\\ Message\\ Types\end{tabular} & \# MQs & \# EQs & \begin{tabular}[c]{@{}c@{}}Time\\ (h:mm)\end{tabular} & Time (ms) &  \\ \hline
Google IoT Core & MQTT V3.1.1 & 115 & 1.00 & 3 & 8s & 5 & 215 & 373 & 06:32 & 0.04 & 06:32 \\ \hline
AWS IoT Core & MQTT V3.1.1 & 102 & 1.00 & 3 & 3s & 5 & 155 & 116 & 02:29 & 0.06 & 02:29 \\ \hline
AWS IoT Core(will) & MQTT V3.1.1 & 103 & 1.00 & 8 & 5s & 4 & 727 & 123 & 04:37 & 0.67 & 04:37 \\ \hline
Azure IoT Hub & MQTT V3.1.1 & 107 & 1.00 & 3 & 8s & 5 & 65 & 393 & 05:31 & 0.04 & 05:31 \\ \hline
Bosch IoT Hub & MQTT V3.1.1 & 106 & 1.00 & 5 & 9s & 5 & 184 & 599 & 09:38 & 0.03 & 09:38 \\ \hline
Aliyun Cloud & MQTT V3.1.1 & 105 & 0.96 & 3 & 4s & 5 & 62 & 1361 & 07:46 & 0.08 & 07:46 \\ \hline
Tuya Smart & MQTT V3.1.1 & 110 & 1.00 & 3 & 8s & 5 & 65 & 393 & 04:53 & 0.03 & 04:53 \\ \hline
Mosquitto & MQTT V5.0 & 106 & 1.00 & 2 & 1s & 5 & 65 & 393 & 00:23 & 0.03 & 00:23 \\ \hline
Mosquitto(will) & MQTT V5.0 & 106 & 1.00 & 6 & 5s & 4 & 317 & 123 & 03:13 & 1.26 & 03:13 \\ \hline
Mosquitto(retain) & MQTT V5.0 & 106 & 1.00 & 8 & 7s & 6 & 727 & 749 & 08:02 & 1.18 & 08:02 \\ \hline
EMQ X & CoAP & 928 & 1.00 & 1 & 1s & 4 & 24 & 420 & 03:47 & 125 & 03:47 \\ \hline
Aliyun Cloud & CoAP & 2152 & 1.00 & 2 & 1s & 3 & 27 & 273 & 04:07 & 1627 & 04:07 \\ \hline
ActiveMQ & AMQP V1.0 & 1808 & 1.00 & 9 & 1s & 8 & 728 & 846 & 05:11 & 1917 & 05:11 \\ \hline
\end{tabular}
\end{table*}

\begin{table*}[htb]
\caption{Performance of \system on property violation detection. \label{tab:performanceproperty}}
\centering
\scriptsize
\begin{tabular}{|c|c|c|c|c|c|c|c|c|c|c|c|}
\hline
\textbf{} & \textbf{\begin{tabular}[c]{@{}c@{}}Google IoT\\ Core\end{tabular}} & \textbf{\begin{tabular}[c]{@{}c@{}}AWS IoT\\ Core\end{tabular}} & \textbf{\begin{tabular}[c]{@{}c@{}}Azure IoT\\ Hub\end{tabular}} & \textbf{\begin{tabular}[c]{@{}c@{}}Bosch IoT\\ Hub\end{tabular}} & \textbf{\begin{tabular}[c]{@{}c@{}}Aliyun\\ Cloud\end{tabular}} & \textbf{Tuya Smart} & \textbf{Mosquitto} & \textbf{EMQ X} & \textbf{\begin{tabular}[c]{@{}c@{}}Aliyun\\ Cloud\end{tabular}} & \textbf{ActiveMQ} & \textbf{Average} \\ \hline
\textbf{Protocol} & \begin{tabular}[c]{@{}c@{}}MQTT\\ v3.1.1\end{tabular} & \begin{tabular}[c]{@{}c@{}}MQTT\\ v3.1.1\end{tabular} & \begin{tabular}[c]{@{}c@{}}MQTT\\ v3.1.1\end{tabular} & \begin{tabular}[c]{@{}c@{}}MQTT\\ v3.1.1\end{tabular} & \begin{tabular}[c]{@{}c@{}}MQTT\\ v3.1.1\end{tabular} & \begin{tabular}[c]{@{}c@{}}MQTT\\ v3.1.1\end{tabular} & \begin{tabular}[c]{@{}c@{}}MQTT\\ v5.0\end{tabular} & CoAP & CoAP & AMQP1.0 & / \\ \hline
\textbf{Precision} & 1.00 & 1.00 & 1.00 & 1.00 & 1.00 & 1.00 & 1.00 & 1.00 & 1.00 & 1.00 & 1.00 \\ \hline
\textbf{\begin{tabular}[c]{@{}c@{}}False Positive rate\end{tabular}}  & 0.00 & 0.00 & 0.00 & 0.00 & 0.00 & 0.00 & 0.00 & 0.00 & 0.00 & 0.00 & 0.00 \\ \hline

\end{tabular}
\end{table*}


\noindent\textbf{Evaluation on property violation detection.} 
Since it is difficult to identify all the security property violations of an MP implementation in practice, we also resort to the 45 experts to manually confirm each of the identified property violations by \system. Therefore, we only report precision. Specifically, the experts act as attackers to perform PoC attacks under the threat models specified in Section~\ref{sec:threatmodel}. 
For secrecy properties, they try their best to retrieve the values of the parameters specified in the target secrecy properties by reversing the traffic, application and device.
If the parameter value can be retrieved, we consider the corresponding secrecy property is violated. As for authentication properties, they try to complete the interactions by forging the messages in the target authentication properties.
If the interactions can be completed by them, we consider the target authentication properties are violated. As a result, the average precision of property violation detection on the ten MP implementations is 100\%.
For more details on the property violation detection, please refer to Table~\ref{tab:performanceproperty}. 

\noindent\textbf{Performance overhead.} We evaluate the overhead of each component in \system and the end-to-end system.
The overall overhead of MP implementations is determinated by the time consumption of the interaction logic extraction module, as other modules' overhead is less than 2152 ms.
The average overhead of the end-to-end system is \textasciitilde4.5 hours.
Considering the interaction logic extraction is a one-shot task, the overhead of \system is acceptable. 
For more details on precision of performance overhead, please refer to Table~\ref{tab:performance}.

\section{Discussions}

\subsection{Lessons}

Based on our evaluation, we conclude that existing popular MPs do not meet the security requirements mainly for the following three reasons. 

\noindent\textbf{Gap between implementations and specifications.}
Many real-world MP implementations do not completely match the standard specification, which on the other hand might be too complex for developers to follow. 
Developers usually have their own understanding about MPs, which leads to some conflicting implementations.
For example, the MQTT on Bosch IoT Hub allows two clients with the same \texttt{ClientID} to be connected to the broker, while the AMQP on ActiveMQ keeps the connection state of a client even when the client is offline.The above implementations all violate their specifications and can be vulnerable.

\noindent\textbf{Gap between constraint resources and security requirements.}
Under the resource-constrained IoT context, developers usually cut down some security functions.
For example, Google IoT Core does not support authentication on the server-side, and the updated version of MQTT on Tuya Smart does not support authentication based on certifications but leverages a vulnerable PSK algorithm instead.
These incomplete security mechanisms are due to that the credential management of numerous devices is challenging and resource-constrained devices cannot support big certificate files.

\noindent\textbf{Gap between the MP security design and adversarial environments.} In terms of the MP design, we find that most developers do not carefully consider the adversarial environments.
First, the adversarial device-sharing cases are not considered. The devices' credentials in some MP implementations are not updated, which may lead to client identity hijacking.
Second, the access control of participants is improper.
For instance, the request/response mechanism introduced by MQTT V5.0 does not limit a client's authority on the response topic, which may cause malicious message injection.

\noindent\textbf{Suggestions.}
With the observations from the security analysis, we make the following suggestions for manufacturers.
First, manufacturers should guarantee secure communications.
The message integrity and confidentiality should be carefully protected. MP implementations should use SSL/TLS with careful configurations, and additional message encryption is highly recommended.
Second, manufacturers need to adopt strict authentication mechanisms. The device and server should not only authenticate the initial connection but also authenticate the messages sent to the agents in every phase.
Besides, the timestamp or message sequences should be applied to avoid replay attacks.
Third, clients' credentials should be dynamically granted to the device or revoked from the device. 
Currently, most MP implementations have hard-coded the device credential into the SDKs, which makes it hard for updating the credentials. Last but not least, the client and server should have fine-grained resource access control.
In particular, we suggest that the identity of a client and her resource should be carefully protected. 

\subsection{Limitations and Future Work}


A limitation of \system is that we only infer the interaction logic and parameter-level semantics of the MP implementations.
An interesting future work is to explore the fine-grained testing and more flexible model learning strategies to catch more fine-grained information of MP implementations.
To illustrate, a bit-wise mutation of a specific parameter in MP messages can help detect if the implementation has appropriately checked the input messages. In addition, it will also be more efficient to apply NLP techniques to analyze the protocol specifications to extract the meta properties.
Also, it is worth mentioning that studying SaaS appliactions might get different results comparing to studying real devices as IoT vendors may configure the SaaS appliactions and introduce some security mechanisms to accomplish the interaction between clients and the server.
 
\section{Related Work}


\noindent \textbf{State machine learning.} A few literature \cite{lie2001simple} works on automatically extracting state machines from protocol implementations.
While these works are effective under the white box setting where the protocol's source code is available, they are not very helpful for MP implementations as most of them are not open-source. In comparison, \texttt{MPInspector} does not use the source code.
Model learning has also been applied to analyze TLS in \cite{de2015protocol}. A similar approach is also used in TLS hostname verification \cite{sivakorn2017hvlearn}. 

\noindent \textbf{Formal verification of protocols.}
In the meanwhile, numbers of verification tools are developed such as ProVerif \cite{blanchet2001efficient} and Tamarin \cite{tamarinmanual}.
Those tools with formal verification have been proved valuable in assessing the security of protocols, such as TLS 1.3 \cite{bhargavan2017verified, cremers2017comprehensive}, LTE \cite{hussain2018lteinspector} and 5G AKA \cite{cremers2019component,basin2018formal}.
By contrast, our framework focuses on the security analysis on protocol implementations.
The idea of combining model learning and model checking was applied in the analysis of TCP and SSH protocols \cite{fiteruau2016combining, fiteruau2017model}.
Comparing to these works, we extend this idea in a more automatic way and come up with the first framework for the security analysis of MP implementations.

\noindent\textbf{Security studies on IoT protocols.}
Researchers have studied the security of IoT communication protocols such as BLE, ZigBee, and Z-Wave \cite{Ronen2017IoT, zwaveattack}. 
However, little work has been done to understand the security of IoT MPs, such as MQTT, AMQP, and CoAP. There are only a few ad-hoc attacks reported.
Previous work \cite{takingmqtt} reveals that attackers can exploit MQTT by connecting the server without authentication and \cite{oconnor2019blinded, zhao2020large} confirmed the attack in real world.
\cite{janes2020never} performed security evaluation on IoT devices' interaction applying the "shared devices attack model".
\cite{oconnor2019homesnitch} presented HomeSnitch to identify a device's behavior in smart home.
In addition, Andrea \emph{et al.} \cite{palmieri2019mqttsa} constructed a tool called MQTTSA to detect the configuration flaw in MQTT deployments based on the source code.
The closest to our work is \cite{jia2020burglars}, which performs a manual security evaluation on MQTT and identifies several design vulnerabilities.
We compare \texttt{MPInspector} with \cite{jia2020burglars} in detail in Section~\ref{sec:comparison}. \texttt{MPInspector} is an automatic approach, covers more MPs and reveals four more new attacks.

\section{Conclusion}

To systematically understand the security of MPs implemented on IoT platforms, we present \texttt{MPInspector}, an automatic and systematic framework to recover MP implementations and reveal the gap between protocol implementations and the desired security properties. \texttt{MPInspector} achieves automated and systematic security analysis by combining model learning and formal analysis. We apply \texttt{MPInspector} to ten implementations of three popular MPs on nine leading commercial IoT platforms, and identify 252 property violations and eleven attacks. We also present the understanding of the MP implementation flaws and discuss the mitigation and future work.
To facilitate future IoT security research, we open source \texttt{MPInspector} at \cite{mpinspector}. 

\section*{Acknowledgments}
We sincerely appreciate our shepherds Omar Chowdhury and Adwait Nadkarni, and all the anonymous reviewers for their valuable comments to improve our paper. We also thank Chenyang Lyu, Yuwei Li, Tianyu Du, Changjiang Li, Yuan Chen, Hong Liang and Han Bao for proofreading this paper.

This work was partly supported by NSFC under No. U1936215, 61772466, and U1836202, the Zhejiang Provincial Natural Science Foundation for Distinguished Young Scholars under No. LR19F020003, the Fundamental Research Funds for the Central Universities (Zhejiang University NGICS Platform), the State Key Laboratory of Information Security (Institute of Information Engineering, Chinese Academy of Sciences, Beijing 100093) (2020-MS-12), the Zhejiang Provincial Natural Science Foundation under No. LQ21F020010, and the Ant Financial Research Funding.


\bibliographystyle{abbrv}
\bibliography{usenix2019_v3.1}

\appendix
\section*{Appendix}

\section{Security properties.}\label{app:properties}
\vspace{-2mm}
We present the main evaluated secrecy and authentication properties (both meta and extended properties) in Table~\ref{tab:secrecyproperties} and Table~\ref{tab:authenticationproperties}, respectively.
\begin{table}[htb]
\caption{Secrecy properties. \label{tab:secrecyproperties}}
\centering
\scriptsize
\begin{threeparttable}
\begin{tabular}{|c|l|}
\hline
\textbf{ID} & \textbf{Secrecy Property Description} \\ \hline
MS1 & Secrecy on MQTT ClientID \\ \hline
$\star$ MS2 & Secrecy on MQTT Secret Key \\ \hline
MS3 & Secrecy on MQTT Username \\ \hline
MS4 & Secrecy on MQTT Password \\ \hline
MS5 & Secrecy on MQTT Topic \\ \hline
MS6 & Secrecy on MQTT Publish Payload \\ \hline
MS7 & Secrecy on MQTT User Properties (MQTT V5.0) \\ \hline
MS8 & Secrecy on MQTT Publish Response Topic (MQTT V5.0) \\ \hline
MS9 & Secrecy on MQTT Publish CorelationData (MQTT V5.0) \\ \hline
AS1 & Secrecy on AMQP ContainerId \\ \hline
AS2 & Secrecy on AMQP Host Name \\ \hline
AS3 & Secrecy on AMQP Transfer Payload \\\hline
AS4 & Secrecy on AMQP Target Node \\\hline
AS5 & Secrecy on AMQP Source Node\\\hline
CS1 & Secrecy on CoAP Uri \\ \hline
CS2 & Secrecy on CoAP Token \\ \hline
CS3 & Secrecy on CoAP MessageId \\ \hline
CS4 & Secrecy on CoAP ACK payload \\ \hline
$\star$ CS5 & Secrecy on CoAP CON\_GET Payload (EMQ X) \\ \hline
$\star$ CS6 & Secrecy on CoAP CON\_PUT Payload (EMQ X) \\ \hline
$\star$ CS7 & Secrecy on CoAP Random (Aliyun Cloud) \\ \hline
$\star$ CS8 & Secrecy on CoAP Secret Key (Aliyun Cloud) \\ \hline
$\star$ CS9 & Secrecy on CoAP AuthToken (Aliyun Cloud) \\ \hline
$\star$ CS10 & Secrecy on CoAP CON\_POSTAUTH payload (Aliyun Cloud) \\ \hline
$\star$ CS11 & Secrecy on CoAP CON\_POSTPUBLISH payload (Aliyun Cloud) \\ \hline
\end{tabular}
\begin{tablenotes}
       \footnotesize
       \item[1] The property with $\star$ is extended property.
       \item[2] MS7-MS9 are only supported in MQTTv5.0, CS5-CS6 are only supported in EMQ X and CS7-CS11 are only supported in Aliyun Cloud in CoAP protocol.
     \end{tablenotes}
   \end{threeparttable}
\end{table}

\begin{table}[hbt!]
\caption{Authentication properties. \label{tab:authenticationproperties}}
\centering
\scriptsize
   \begin{threeparttable}
\begin{tabular}{|c|l|}
\hline
\textbf{ID} & \textbf{Property Description} \\ \hline
MA1 & Authentication on MQTT CONNECT message (server-\textgreater{}client) \\ \hline
MA2 & Authentication on MQTT CONNACK message (client-\textgreater{}server) \\ \hline
MA3 & Authentication on MQTT SUBSCRIBE message (server-\textgreater{}client) \\ \hline
MA4 & Authentication on MQTT SUBACK message (client-\textgreater{}server) \\ \hline
MA5 & Authentication on MQTT UNSUBSCRIBE message (server-\textgreater{}client) \\ \hline
MA6 & Authentication on MQTT UNSUBACK message (client-\textgreater{}server) \\ \hline
MA7 & Authentication on MQTT PUBLISH message (server-\textgreater{}client) \\ \hline
MA8 & Authentication on MQTT PUBACK message (client-\textgreater{}server) \\ \hline
MA9 & Authentication on MQTT DISCONNECT message (server-\textgreater{}client) \\ \hline
MA10 & \begin{tabular}[c]{@{}l@{}}Authentication on MQTT Will message PUBLISH message \end{tabular} \\ \hline
MA11 & \begin{tabular}[c]{@{}l@{}}Authentication on MQTT Retained message PUBLISH message \end{tabular} \\ \hline
AA1 & Authentication on AMQP SASL message (server-\textgreater{}client) \\ \hline
AA2 & Authentication on AMQP SASL message (client-\textgreater{}server) \\ \hline
AA3 & Authentication on AMQP OPEN message (server-\textgreater{}client) \\ \hline
AA4 & Authentication on AMQP OPEN message (client-\textgreater{}server) \\ \hline
AA5 & Authentication on AMQP ATTACH message (server-\textgreater{}client) \\ \hline
AA6 & Authentication on AMQP ATTACH message (client-\textgreater{}server) \\ \hline
AA7 & Authentication on AMQP FLOW message (server-\textgreater{}client) \\ \hline
AA8 & Authentication on AMQP FLOW message (client-\textgreater{}server) \\ \hline
AA9 & Authentication on AMQP TRANSFER message (server-\textgreater{}client) \\ \hline
AA10 & Authentication on AMQP DISPOSITION message (client-\textgreater{}server) \\ \hline
AA11 & Authentication on AMQP DETACH message (server-\textgreater{}client) \\ \hline
AA12 & Authentication on AMQP DETACH message (client-\textgreater{}server) \\ \hline
AA13 & Authentication on AMQP CLOSE message (server-\textgreater{}client) \\ \hline
$\star$ CA1 & Authentication on CoAP CON\_GET message (EMQ X) (server-\textgreater{}client) \\ \hline
$\star$ CA2 & Authentication on CoAP CON\_GET message (EMQ X) (client-\textgreater{}server) \\ \hline
$\star$CA3 & Authentication on CoAP CON\_PUT message (EMQ X) (server-\textgreater{}client) \\ \hline
$\star$ CA4 & Authentication on CoAP CON\_PUT message (EMQ X) (client-\textgreater{}server) \\ \hline
$\star$ CA5 & \begin{tabular}[c]{@{}l@{}}Authentication on CoAP CON\_POSTAUTH message (Aliyun Cloud)\\ (server-\textgreater{}client)\end{tabular} \\ \hline
$\star$ CA6 & \begin{tabular}[c]{@{}l@{}}Authentication on CoAP CON\_POSTAUTH message (Aliyun Cloud))\\ (client-\textgreater{}server)\end{tabular} \\ \hline
$\star$ CA7 & \begin{tabular}[c]{@{}l@{}}Authentication on CoAP CON\_POSTPUBLISH message (Aliyun Cloud)\\ (server-\textgreater{}client)\end{tabular} \\ \hline
$\star$ CA8 & \begin{tabular}[c]{@{}l@{}}Authentication on CoAP CON\_POSTPUBLISH message (Aliyun Cloud)\\ (client-\textgreater{}server)\end{tabular} \\ \hline
\end{tabular}
\begin{tablenotes}
       \footnotesize
       \item[1] The property with $\star$ is extended property.
       \item[2] Authentication properties on both client side and server sides are considered. CA1-CA4 are only supported in EMQ X and CA6-CA7 are only supported by Aliyun Cloud in CoAP protocols.
      \item[3] A-\textgreater B means that A authenticates the message from B.
     \end{tablenotes}
\end{threeparttable}
\end{table}

\vspace{-3mm}
\section{A Running Example}
\label{app:arunningexample}
\vspace{-2mm}

We take the MQTT implementation on Bosch IoT platform as a running example to clarify how the state machine is generated and how the formal code is translated.

\noindent\textbf{State machine and property generation.} 
First, \system applies message semantics extraction from Section~\ref{sec:parametersemanticsextraction} to identify the parameter semantics for the key messages specified in the MQTT standard.
In particular, \texttt{MPInspector} outputs the semantics of nine key MQTT messages using the JSON encoding, e.g., \code{\{"CONNECT":\{"ClientID":"","username":\{"composition":["authid","tenantid"]\},"password":""\}\}} (an expression \texttt{"parameter":""} means that \texttt{parameter} does not have extra semantics and is consistent with the standard MP).

\begin{figure}[htbp]
    \centering
    \includegraphics[width=0.50\textwidth]{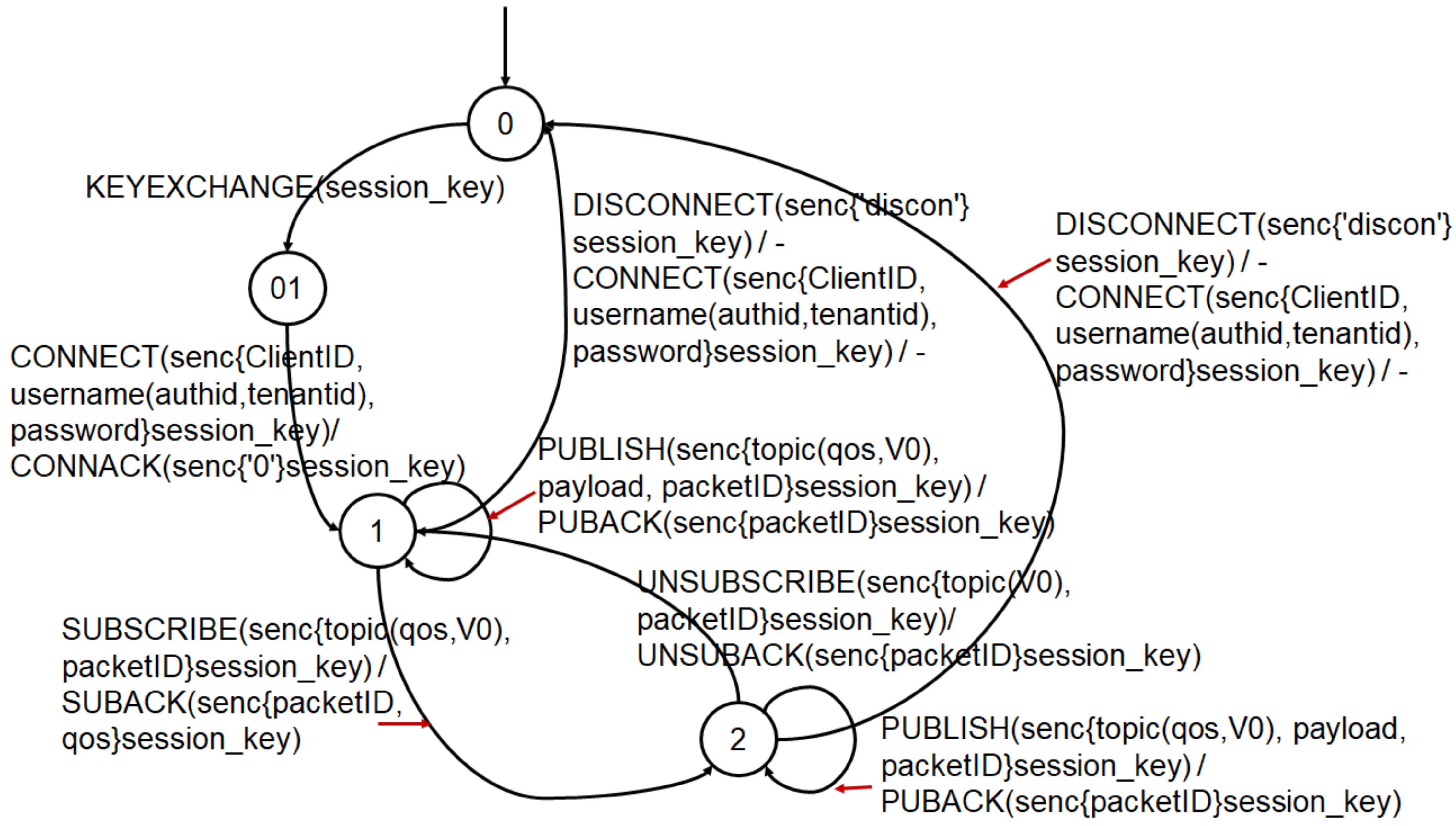}
    \caption{The inferred state machine of the MQTT implementation on the Bosch IoT platform.}
    \label{fig:bosch}
\end{figure}

Second, \texttt{MPInspector} applies interaction logic extraction from Section~\ref{sec:parametersemanticsextraction} to the MQTT implementation on the Bosch IoT platform.
It outputs a raw state machine whose transition messages only contains the message names, e.g., \texttt{CONNECT/CONNACK}.
Then, it adds the semantics extracted from Section~\ref{sec:parametersemanticsextraction}  to each transition message.
After that, we have the inferred state machine as shown in Figure~\ref{fig:bosch}.
According to the property generation method in  Section~\ref{sec:propertiesgeneration}, \texttt{MPInspector} outputs the secrecy and authentication properties as shown in Appendix~\ref{app:properties}.

\noindent\textbf{State machine translation.}
First, \texttt{MPInspector} generates the special initial rule and session key negotiation rule.
The initial rule defines the initial states of the broker and clients, which is concluded from the MQTT specification.
\texttt{MPInspector} uses the let-binding expression to specify the parameter semantics in the initial states, as shown in the second line of Listing~\ref{lst:initrule}.
\texttt{MPInspector} generates the transition rule for session key negotiation based on the state machine, which is a simplified SSL/TLS key negotiation modeling.
The rule is shown in Listing~\ref{lst:keyrule}. 

\begin{lstlisting}[breaklines,basicstyle=\footnotesize,caption={An example of an initial rule in Tamarin code.},label={lst:initrule}, frame = single]
rule init_client:
    let username = <~authid,~tenantid> in
    [ !SERVER($SERVER), Fr(~ClientID), Fr(~authid),Fr(~tenantid), Fr(~password)]--[  ]->[!DEVICE($SERVER,~ClientID,username,~password),!State_0_Serv($SERVER,~ClientID,username,~password),!State_0_Dev($SERVER,~ClientID,username,~password)]
\end{lstlisting} 

\begin{lstlisting}[breaklines,basicstyle=\footnotesize,caption={An example of a session key negotiation rule in Tamarin code.},label={lst:keyrule}, frame = single]
rule client_serv_negotiate_tls_key:
    let username = <~authid,~tenantid> in 
    [!State_0_Serv($SERVER,~ClientID,username,~password),!State_0_Client($SERVER,~ClientID,username,~password),Fr(~session_key)]
  --[  ]->[ Dev_Tls_Sym($SERVER,~ClientID,username,~password,~session_key),
    Serv_Tls_Sym($SERVER,~ClientID, username, ~password,~session_key)]
\end{lstlisting}

Second, we translate the transition messages from the inferred state machine to rules following the principle described in Section~\ref{sec:formalcodetranslation}.
Taking the server side transition \code{CONNECT(senc\{ClientID,username(V1,V2),password\}session\_key/CONNACK(senc('0')session\_key))} as an example, we show its translated Tamarin rule in Listing~\ref{lst:connectrule}.
As shown in Listing~\ref{lst:connectrule}, the rule's left part shows the state that the server receives the \texttt{CONNECT} message and its right part indicates the state that the server sends out the \texttt{CONNACK} message.
The action facts in the rule's middle part indicate the behaviors in the transition, which will be used in the property lemmas for reasoning.
For example, \texttt{Secret(<'server','password',password>)} means that the \texttt{password} is supposed to be secret on the server side.

\begin{lstlisting}[breaklines,basicstyle=\footnotesize,caption={An example of a transition rule in Tamarin code.},label={lst:connectrule}, frame = single]
rule serv_recv_connect_snd_conncak:
  let username = <authid,tenantid>
    uername =  <authid,tenantid>
    connack = senc('0',session_key)
    connect = senc{ClientID,username,password}session_key
  in [ In(connect), Serv_Tls_Sym($SERVER,ClientID,authid, tenantid,password,session_key) ] --[ Create('connect','server',$SERVER), Commit($SERVER,username,<'server','client',username>), Commit($SERVER,username,<'server','client',ClientID>), Commit($SERVER,username,<'server','client',password>), Running($SERVER,username,<'client','server',<'connack',connack>>), Honest(<'client',username>),Honest(<'server',$SERVER>), Secret(<'server','username',username>), Secret(<'server','password',password>), Secret(<'server','ClientID',ClientID>)] ->[ Out(connack), State_1_Serv($SERVER,ClientID,authid,tenantid,password,session_key]
\end{lstlisting}

\noindent\textbf{Property translation.} 
Finally, the formal code translation module automatically translates the secrecy properties on password to Tamarin code using the formula shown in Listing~\ref{lst:seclemma}.
\texttt{MPInspector} automatically generates four types of authentication lemmas for each authentication property based on the state machine.
Taking the injective agreement as an example, \system generates the formalization of the injective agreement property on a \texttt{CONNECT} message, as shown in Listing~\ref{lst:authlemma}.
Listing~\ref{lst:seclemma} and Listing~\ref{lst:authlemma} show the property lemmas use the first-order logic formulas over time points and acton facts, based on the standard security property templates specified by Tamarin Prover \cite{tamarinmanual}.

\begin{lstlisting}[breaklines,basicstyle=\footnotesize,caption={An example of a secrecy lemma in Tamarin code.},label={lst:seclemma}, frame = single]
lemma secret_Password_serv:
"All n #i. Secret(<'server','password',n>) @i ==> (not (Ex #j. K(n)@j)) | (Ex A B #j. Reveal(A,B)@j & Honest(A)@i)"
\end{lstlisting}

\begin{lstlisting}[breaklines,basicstyle=\footnotesize,caption={An example of an authentication lemma in Tamarin code.},label={lst:authlemma}, frame = single]
lemma injective_agreement_dev_serv_CONNECT:   
  "All a b t #i. Commit(a,b,<'server','client',t>) @i ==> (Ex #j. Running(b,a,<'server','client',t>) @j & j < i & not (Ex a2 b2 #i2. Commit(a2,b2,<'server','client',t>) @i2 & not (#i2 = #i))) | (Ex C data #r. Reveal(C,data)@r & Honest(C) @i)"
\end{lstlisting}


\end{document}